%% file: paper1.tex
\documentclass{jpp}

\usepackage{graphicx}
\usepackage{natbib}
\usepackage[colorlinks,urlcolor=blue,citecolor=blue,linkcolor=blue]{hyperref}
\usepackage{amsmath}

\ifCUPmtlplainloaded \else
  \checkfont{eurm10}
  \iffontfound
    \IfFileExists{upmath.sty}
      {\typeout{^^JFound AMS Euler Roman fonts on the system,
                   using the 'upmath' package.^^J}%
       \usepackage{upmath}}
      {\typeout{^^JFound AMS Euler Roman fonts on the system, but you
                   dont seem to have the}%
       \typeout{'upmath' package installed. JPP.cls can take advantage
                 of these fonts, if you use 'upmath' package.^^J}%
      }
  \else
  \fi
\fi

\ifCUPmtlplainloaded \else
  \checkfont{msam10}
  \iffontfound
    \IfFileExists{amssymb.sty}
      {\typeout{^^JFound AMS Symbol fonts on the system, using the
                'amssymb' package.^^J}%
       \usepackage{amssymb}%
       \let\le=\leqslant  \let\leq=\leqslant
         
      }{}
  \fi
\fi

\ifCUPmtlplainloaded \else
  \IfFileExists{amsbsy.sty}
    {\typeout{^^JFound the 'amsbsy' package on the system, using it.^^J}%
     \usepackage{amsbsy}}
    {\providecommand\boldsymbol[1]{\mbox{\boldmath $##1$}}}
\fi

\newcommand{\bmath}[1]{{\bf {#1}}}

\newcommand{\bJ}{\bmath{J}}

\newcommand{\bE}{\bmath{E}}

\newcommand{\vpr}[2]{\bmath{#1} \!\times\! \bmath{#2}}

\newcommand{\Od}[1]{\frac{d}{d #1}}

\newcommand{\fracp}[2]{\left(\frac{#1}{#2}\right)}

\newcommand{\eq}[1]{Eq.~(\ref{eq:#1})}

\def\comp{\,c/\omega_{\rm p}}

\newcommand{\fig}[1]{Fig.~\ref{fig:#1}}
\newcommand{\gammamax}{\gamma_{\rm max}}

\newcommand{\be}{\begin{equation}} 
\newcommand{\ee}{\end{equation}}
\newcommand{\nn}{\mbox{} \nonumber \\ \mbox{} }
\newcommand{\ba}{\begin{eqnarray}}
\newcommand{\ea}{\end{eqnarray}}
\newcommand{\om}{\omega}
\newcommand{\Alfven}{Alfv\'{e}n }

\newcommand{\Bf}{{magnetic field}}

\newcommand{\Ef}{{electric  field}}
\newcommand{\Efs}{{electric fields}}

\newcommand{\EM}{electromagnetic}

\newcommand{\curl}{{\rm curl\, }}
\newcommand{\E}{{\bf E}}
\newcommand{\B}{{\bf B}}

\newcommand{\ex}[1]{\times10^{#1}}

\newcommand{\aap}{    {\it Astron. Astrophys.}}
\newcommand{\aaps}{   {\it Astron. Astrophys. Suppl.}}
\newcommand{\aapr}{   {\it Astron. Astrophys. Rev.}}

\newcommand{\aj}{     {\it Astron. J.}}
\newcommand{\apj}{    {\it Astrophys. J.}}
\newcommand{\apjl}{    {\it Astrophys. J. Lett.}}
\newcommand{\apss}{   {\it Astrophys. Space Sci.}}

\newcommand{\fcp}{    {\it Fundamenals Cosm. Phys.}}

\newcommand{\grl}{    {\it Geophys. Res. Lett.}}

\newcommand{\jgr}{    {\it J. Geophys. Res.}}
\newcommand{\mnras}{  {\it Mon. Not. Roy. Astron. Soc.}}
\newcommand{\nat}{    {\it Nature}}
\newcommand{\pasp}{   {\it Pub. Astron. Soc. Pac.}}
\newcommand{\pasj}{   {\it Pub. Astron. Soc. Japan}}
\newcommand{\prd}{    {\it Phys. Rev. D}}
\newcommand{\pre}{    {\it Phys. Rev. E}}
\newcommand{\solphys}{{\it Solar Phys.}}
\newcommand{\sovast}{ {\it Sov. Astron.}}
\newcommand{\ssr}{    {\it Space Sci. Rev.}}

\newcommand{\na}{    {\it Nature}}
\def\physrep{Phys.~Rep.}      % Physics Reports
\def\araa{ARA\&A}             % Annual Review of Astron and Astrophys

\title[Explosive  $X$-point collapse]{Explosive  $X$-point collapse in relativistic  magnetically-dominated plasma}

\author{Maxim Lyutikov,$^1$  Lorenzo Sironi$^{2}$, Serguei S. Komissarov$^{1,3}$,  Oliver Porth$^{3,4}$}
\affiliation{$^1$ Department of Physics, Purdue University, 
 525 Northwestern Avenue,
West Lafayette, IN
47907-2036, USA; lyutikov@purdue.edu
\\
$^2$  Department of Astronomy, Columbia University, 550 W 120th St, New York, NY 10027, USA; lsironi@astro.columbia.edu
\\
$^3$  School of Mathematics,
University of Leeds, LS29JT
Leeds, UK; s.s.komissarov@leeds.ac.uk
\\
$^4$  Institut f\"{u}r Theoretische Physik, 
J. W. Goethe-Universit\"{a}t, 
D-60438, Frankfurt am Main, Germany
porth@th.physik.uni-frankfurt.de
}

\begin{document}

\maketitle

%%%%%%%%%%%
\input{abstract1.tex}

\clearpage
%%%%%%%%%

%\tableofcontents

\input{intro1.tex}

%%%%%%%Maxim%%%%%
\input{X-point.tex}

%\clearpage
%%%%%%%%%%%%
%sssssssssssssssssssssssssssssssssssssssssssss
%\section{$X$-point collapse: numerical simulations}
%\label{sims}

%%%%%%%Sergey%%%%%%%
%\subsection{Force-free simulations: X-point collapse}
\input{FF-X.tex}

%\clearpage
%%%%%%%

%%%%%%%Lorenzo%%%%%%%
\input{PIC1.tex}

%\clearpage
%%%%%%%

\input{conclusion1.tex}

\acknowledgements

We would like  to thank Roger Blandford,  Krzystof Nalewajko,  Dmitri Uzdensky and Jonathan Zrake for numerous and helpful discussions.

The PIC simulations were performed on XSEDE resources under contract No. TG-AST120010, and on NASA High-End Computing (HEC) resources through the NASA Advanced Supercomputing (NAS) Division at Ames Research Center. ML would like to thank for hospitality Osservatorio Astrofisico di Arcetri and Institut de Ciencies de l'Espai, where large parts of this work were conducted. SSK thanks Purdue University for hospitality during his sabbatical in 2014. 
This work had been supported by NASA grant NNX12AF92G,  NSF  grant AST-1306672 and DoE grant DE-SC0016369.
OP is supported by the ERC Synergy Grant ``BlackHoleCam -- Imaging the Event Horizon of Black Holes'' (Grant 610058).
SSK is supported by the STFC grant ST/N000676/1. 
  
\clearpage
\bibliographystyle{jpp}
%\bibliographystyle{mn2e}
%  \bibliography{astro}
 \bibliography{/Users/maxim/Home/Research/BibTex} 
\appendix

%%%%%%%%%%%%%%%%%
\input{appendix1.tex}

\clearpage
%%%%%%%%%
%%%%%

   \end{document}

%% file: abstract1.tex
The extreme properties of the gamma ray flares in the Crab Nebula present a clear challenge to our ideas on the nature of particle acceleration in relativistic astrophysical plasma. It seems highly unlikely that standard mechanisms of stochastic type are at work here and hence the attention of theorists has switched to linear acceleration in magnetic reconnection  events. In this series of papers, we attempt to develop a theory of explosive magnetic reconnection in highly-magnetised  relativistic plasma which can explain the extreme parameters of the Crab flares.    In the first paper, we focus on the properties of the X-point collapse.  Using analytical and numerical methods (fluid and particle-in-cell simulations)  we extend Syrovatsky's classical model of  such collapse to the relativistic  regime. We find that the collapse can lead 
to the reconnection rate approaching the speed of light on macroscopic scales.
%The fastest particle  acceleration occurs during the initial  stages of the catastrophic  X-point collapse.  
During the collapse, the plasma particles are accelerated by charge-starved \Efs, which can reach (and even exceed) values of  the local \Bf. The explosive stage of reconnection produces non-thermal power-law tails 
with slopes that depend on the average magnetization $\sigma$.  For sufficiently high magnetizations and vanishing guide field, the non-thermal particle spectrum consists of  two components: a low-energy population with soft spectrum, that dominates the number census; and a high-energy population with hard spectrum, that possesses all the properties needed to explain the Crab flares.

%% file: intro1.tex
%sssssssssssssssssssssssssssssssssssssssssssss
\section{Introduction}
\label{intro}
%sssssssssssssssssssssssssssssssssssssssssssss

The detection of  flares from  Crab Nebula by AGILE and Fermi satellites  \citep{2011Sci...331..736T,2011Sci...331..739A,2012ApJ...749...26B} is one of the most astounding discoveries in high energy astrophysics. 
The unusually short durations, high luminosities, and high photon energies of the Crab Nebula gamma-ray flares require reconsideration of our basic assumptions about the physical processes responsible for acceleration of the highest-energy emitting particles in the Crab Nebula, and, possibly in other high-energy  astrophysical sources. 

The Crab flares are characterised by an increase of  gamma-ray flux above 100 MeV by a factor of few or more on 
the day time-scale.  This energy corresponds to the high end of the Crab's synchrotron spectrum.   
Most interestingly,  in the other energy bands nothing unusual has been observed  during the flares so far \citep{2013ApJ...765...56W}. This suggests that the physical processes behind the flares lead to a dramatic increase of the highest energy population of relativistic electrons in the nebula, whereas lower energy population remains largely unaffected.  The short duration of flares indicate explosive and highly localised events. 
   
Most importantly, the peak of the flare spectrum approaches and even exceeds the maximal rest-frame synchrotron photon energy  \citep{1996ApJ...457..253D,2010MNRAS.405.1809L,2012MNRAS.426.1374C}. 
Balancing the synchrotron energy losses in the magnetic field $B$ against     
the energy gain via acceleration in the electric field of strength $E=\eta B$ leads to the upper limit 
of the synchrotron photon energy 
$$
\epsilon_{\rm max} \sim \eta   \hbar  { m c^3 \over e^2} \approx 100 \eta \mbox{ MeV}
\label{emax}
$$
The high conductivity of astrophysical plasma ensures that for typical accelerating electric field  $\eta < 1$.  The fact that the flare spectrum extends beyond this limit pushes $\eta$ towards unity, which implies energy gain on the scale of the gyration period.   This  practically {\it excludes stochastic acceleration mechanisms} in general and the shock acceleration in particular. In principle, strong Doppler boosting could somewhat reduce this constraint \cite{2011MNRAS.414.2229B,2012MNRAS.426.1374C} but the  lack of observational evidence for ultra-relativistic macroscopic motion inside the nebula  makes this unlikely.

 A widely discussed alternative to the shock acceleration mechanism is the particle acceleration accompanying  
 magnetic reconnection.  It is well known that magnetic reconnection can lead to explosive release of magnetic energy, e.g. in solar flares.  However, properties of plasma in the Crab Nebula, as well as magnetospheres of pulsars and magnetars, pulsar winds, AGN and GRB jets and other targets of relativistic astrophysics,  are very different from those of more conventional Solar and laboratory plasmas \citep{2013SSRv..178..459L}.  In particular, the energy density of magnetic field can exceed not only the thermal energy density but also the rest mass-energy density of plasma particles.  In order to quantify such a strong magnetization, it is convenient to use the relativistic magnetization parameter 
\be
\sigma =  \frac{B^2}{4 \pi w}
\ee
where $w=\rho c^2 + (\hat{\gamma}p /\hat{\gamma}-1)$ is the relativistic enthalpy, which includes the rest mass-energy density of plasma. In traditional plasmas this parameter is very small but in relativistic astrophysics
$\sigma \gg 1$ is quite common.  This parameter is uniquely related to the \Alfven speed $v_A$  via 
$$ 
(v_A/c)^2 = \sigma /(1+\sigma) \,.
$$
   
The physics of particle acceleration in relativistic current sheets has been addressed in a number of recent studies \citep[\eg][and others]{2012ApJ...750..129B,2015ApJ...806..167G,nalewajko_15,2015ApJ...805..163D}. In particular, the
 Colorado group \citep{2011ApJ...737L..40U,2012ApJ...746..148C,2012ApJ...754L..33C,2013ApJ...770..147C} explored the possibility of explaining  the Crab flares. Their 2D PIC simulations demonstrated the development of the 
relativistic tearing instability  \citep[see also][]{LyutikovTear,2007MNRAS.374..415K}, followed by a transition to the plasmoid-dominated regime.   In addition to a number of useful properties, the current sheet reconnection has some generic features  which seem to be in conflict with the observations of the Crab flares.  

First, in the collisionless plasma the key length scales of reconnection current sheet are microscopic and determined by to the plasma skin depth.   This applies to the thickness of the current sheet, the distance between plasmoids (magnetic ropes) and the correlation scale of the accelerating electric  field.  As a result the typical potential available for linear 
acceleration is limited to that over microscopic scales (few skin depths), which is too small to explain the flares spectrum. The correlation
scale can be increased with introduction of strong guide field but this leads to a significant reduction of the reconnection rate  \citep{zenitani_08}. 
% {\bf SSK: We need some supporting estimates here. ML: I think we are OK as is. This is a fairly well established fact by now, so we do not want to spend time describing it.}

Second, the recent PIC studies of relativistic reconnection have demonstrated that even in the absence of the guide field the reconnection rate (inflow velocity) in 3D simulations is significantly lower than in 2D \citep{2014ApJ...783L..21S}. Thus, for a reference magnetization $\sigma=10$  the reconnection rate in 2D is $r_{\rm rec}\sim 0.1$ whereas in 3D it is only 
$r_{\rm rec}\sim 0.02$  \citep{2014ApJ...783L..21S}. The slower reconnection rate leads to a weaker accelerating electric field. Moreover, for a given flare duration it  translates into a smaller utilised magnetic energy.  
%Also, the development of the tearing mode generally does not lead to  global reconfigurations - it  typically results in a local dissipation event  \citep[like saw-tooth oscillations in TOKAMAKs][]{1987RvMP...59..175F}. 

Finally, and most importantly, Crab flares require acceleration to Lorentz factors $\gamma \sim 10^9$. In the simulations of  \citep{2011ApJ...737L..40U,2012ApJ...746..148C,2012ApJ...754L..33C,2013ApJ...770..147C} {\it  all} the particles present within the acceleration region get accelerated  to similar energies. In such a case, the terminal Lorentz factor is  limited by $\gamma_{max} \sim \sigma$, which requires unreasonably highly magnetized regions to exist inside the Crab nebula.  

These problems may stem from the simplified slab (or, plane-parallel) geometry of the initial configuration, enforced by the
periodic boundary conditions, which was considered in the above studies. Indeed, this excludes the large-scale magnetic stresses which in highly-magnetized plasma may lead to explosive high-speed dynamics on macroscopic length scales.  
In this series of papers, we explore the role of the macroscopic factor by considering various initial configurations and studying their macroscopic evolution using fluid-type models of magnetized relativistic plasma. Each such study is accompanied by PIC simulations, which allows to study the specifics of particle acceleration resulted from the involved magnetic reconnection.  We start by considering the classical problem of the X-point collapse.    

The theoretical studies of the X-point collapse trace back to the work by \cite{doi:10.1080/14786440708521050}, who argued that a neutral $X$-point is unstable and that particles can be accelerated during its collapse. The ideas of Dungey were put on a firm basis by \cite{1967JETP...25..656I}, who found corresponding non-linear MHD solutions  \citep[see also][]{2000mare.book.....P}. The solutions of \cite{1969ARA&A...7..149S,1967JETP...25..656I,1981ARA&A..19..163S}  were 
obtained in what we can nowadays call a quasi-static force-free approximation: neglecting the pressure effects (hence force-free), but also neglecting the dynamic effects of the electric field (hence the name quasi-static). 
In this paper we extend these studies by considering the case of highly-magnetized plasma ($\sigma\gg 1$). 

We start by describing an approximate analytical solution found in the framework of force-free electrodynamics (Section~\ref{X-point}). The solution shows that in highly-magnetised plasma X-points remain unstable to collapse. This result is verified by our numerical simulations, which also allow a more comprehensive study the dynamics of X-point collapse (Sections~\ref{FF-X} and \ref{PIC1}). Using 2D Particle-in-Cell (PIC) simulations, we studied the process of non-thermal particle   
acceleration accompanying the collapse (Sections~\ref{PIC1}). The results show a number features, which have not been seen in previous studies, focused on relativistic reconnection in plane current sheet, and which may have important implications in the theory of Crab flares.  In Section~\ref{Conc1} we summarise our main results and discuss their implications.

%% file: X-point.tex
%sssssssssssssssssssssssssssssssssssssssssssss
 \section{Asymptotic model of X-point collapse in force-free plasma}
\label{X-point}

\subsection{Dynamic force-free plasma}

Explosive release of magnetic energy is a common phenomenon in laboratory and space plasmas \citep[\eg][]{2000mare.book.....P}.  Syrovatskii \citep{1967JETP...25..656I,1975IzSSR..39R.359S,1981ARA&A..19..163S}  argued that it could be related to the macroscopic instability of magnetic X-point configuration and studied it in the framework of non-relativistic  MHD   \citep[see also][]{1997PhR...283..185C}. In this Section, we develop this theory farther by considering the case  of highly magnetized relativistic plasma with $\sigma \gg 1$. In this regime, (i) the mass-density of 
plasma is dominated by the magnetic field, (ii) the \Alfven speed approaches the speed of light, (iii) the conduction current flows mostly along the \Bf lines, (iv) the displacement current $(c / 4 \pi) \partial_t {\bf E}$ may be comparable to the conduction current, ${\bf j}$, (v) the electric charge density, $\rho_e$, may be of the order of $j/c$.  

The large value of $\sigma$ (or small $1/\sigma$) can be used as an
expansion parameter in the equations of relativistic magnetohydrodynamics. The zero order equations describe 
the so-called {\it force-free electrodynamics}  or {\it magnetodynamics}, \cite{2002MNRAS.336..759K}. 
In this approximation, the inertia of plasma particle is ignored and hence the macroscopic Lorentz
force completely vanishes. Hence the energy and momentum of the electromagnetic fields are conserved.    
The perfect conductivity condition reduces to ${\bf E} \cdot {\bf B}=0$ and $E^2<B^2$, which 
ensure existence of inertial frames where $\bE=0$. 
   
The electric current of ideal force-free electrodynamics  can be written entirely in terms of the electromagnetic field and its spatial derivatives   
\be
{\bf J}= {c \over 4 \pi} {({\bf E}\times{\bf B})\nabla\cdot{\bf E}+
({\bf B}\cdot\nabla\times{\bf B}-{\bf E}\cdot
\nabla\times{\bf E}){\bf B}\over   B^2}
\label{FF}
\ee
\citep{1997PhRvE..56.2181U,Gruzinov99}. This may be considered as the Ohm's law of this approximation.  Similarly to resistive MHD, one can modify this law and introduce magnetic dissipation \citep[\eg][see also \S \ref{FF-X}]{LyutikovTear,2012ApJ...746...60L}. Importantly, only the parallel component of the electric current is subject to resistive dissipation.

\subsection{Stressed X-point collapse in force-free plasma: analytical solution}

The non-current-carrying unstressed X-point configuration with translational symmetry along the $z$ axis is described by the vector potential 
$$
A_z =-  \left( {x^2  - y^2} \right) { B_0 \over 2 L}\, .
$$ 
It has null lines intersecting at 90 degrees at the origin, where the magnetic field vanishes. $B_0$ is the magnetic field strength at the distance $L$ from the origin.  When uniformly squeezed, such an X-point is known to be unstable to collapse in the non-relativistic regime \citep{1953MNRAS.113..180D,1967JETP...25..656I,2000mare.book.....P}.  
In the force-free limit, such a solution develops a singularity from the start, unless we introduce a guide field. 
 For simplicity, here we consider a uniform guide field $B_z=B_0$, which makes $L$  a characteristic length scale of our problem: at $r=L$ the guide field has the same strength as the field of the X-point. 
In the following, we deal with dimensionless equations using $L$ as the unit length scale,  $L/c$ as the unit time scale 
and  $B_0$ as the unit field strength.  

Following the previous work on the X-point collapse,  we look for approximate solutions in the form
\be
A_z =- \frac{1}{2} \left( {x^2 \over a(t)^2} -{y^2\over b(t)^2} \right) \, ,
\ee
which assumes that at all times the configuration can be described as uniformly squeezed. This configuration 
is similar to the unstressed one but the null lines run at an angle determined by the ratio $a/b$.  Without loss of generality, we 
can put $a(0)=1$ and $b(0)=\lambda$, making $\lambda$ a parameter describing the initial perturbation.   
Such solutions exists only in the limit $x,y\ll 1$ and $t \ll 1$, where the guide field remains largely unchanged.  
Hence we put   
\ba &&
\B = \curl (A_z {\bf e}_z)+ {\bf e}_z
\nn &&
\E  =- \partial_t {\bf A} + \nabla \Phi(t, x,y)
\ea
and the force-free condition $\E \cdot \B=0$ reads
\be
\left( - x^2 {\partial _t a \over a^3} +  y^2 {\partial _t b \over b^3} \right) + 
\left(  {x \partial_y \Phi \over a^2} +{ y \partial_x \Phi \over b^2} \right)  =0 \,.
\ee
This implies  
\ba &&
b(t) = \lambda /a(t)
\nn &&
\Phi = xy\, \frac{ \dot{a}}{a}
\ea
and hence  $a(t)$ (or $b(t)$) is the only unknown function of the problem\footnote{Notice that  $\Delta \Phi=0$ and hence the electric charge density remains vanishing all the time.}. 
Substituting the expressions for the electric and magnetic fields into the Maxwell equation yields  an ordinary differential equation for $a(t)$.  Since $x,y \ll 1$ 

$$ 
\B \cdot \curl \B \simeq ( 1/a^2 - (a/\lambda)^2) )\,, 
$$
$$
\E \cdot \curl \E \simeq   2 \dot{a}^2  
\left( \frac{1}{a^4} {x}^2+  \frac{1}{\lambda^2} {y}^2 \right)\,,
$$
and the Maxwell  equation reduces to 
\be
\Od{t} \fracp{\dot{a}}{a}    = { a^4-\lambda^2 \over \lambda^4}  \,.
\label{addot}
 \ee
Given $a(t)$ the \EM\ field can be found via 
\ba &&
\B= \left( { y a^2 \over \lambda^2} ,  {x\over a^2}, 1 \right) \,,
\nn &&
\E=  \left(  y,  x , - { x^2 \lambda^2 + y^2 a^4 \over \lambda^2 a^2} \right)  \frac{\dot{a}}{a} \,.
\label{eq:coll}
\ea

\begin{figure}
\centering
\includegraphics[width=.99\textwidth]{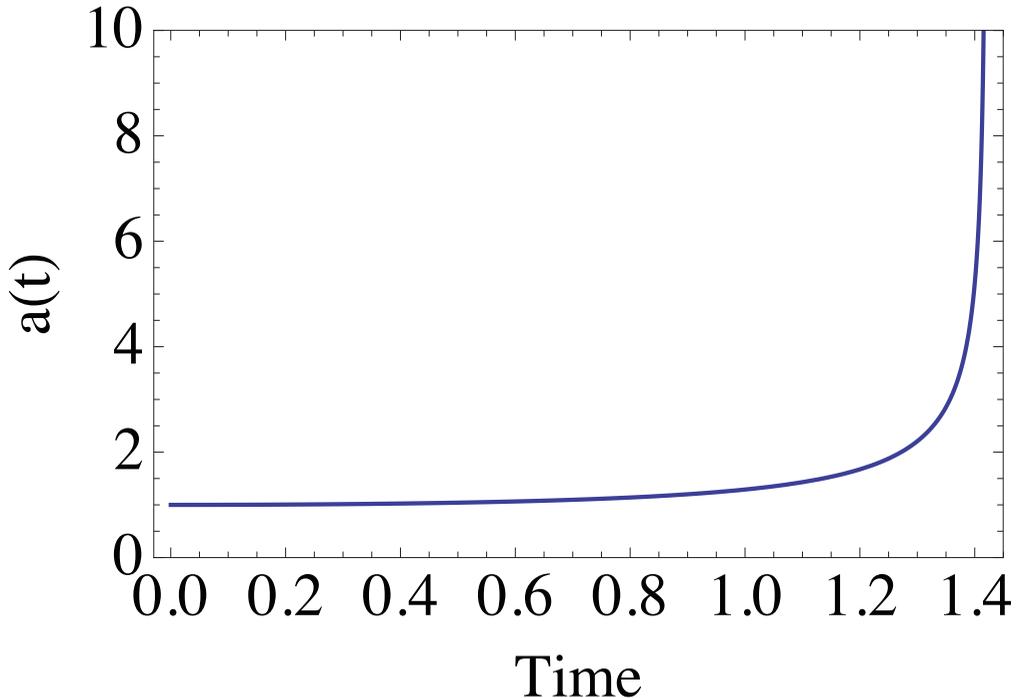}
\caption{ An example of the evolution of the parameter $a(t)$, Eq. (\ref{addot}). Initially an X-point is squeezed by ten percent, $\lambda =0.9$,  parameter ${\cal A}=1$. Evolution occurs on the dynamical time scale, until a singularity  at  $t=1.42$, so that   the fast growing  stage of the collapse proceeds much quicker. }
\label{aoft} 
\end{figure}

 Solutions of the equations  (\ref{addot}) show that ``the sqeezinees''  parameter $a(t)$ has  a finite time singularity for $\lambda < 1$: in finite time $a$ becomes infinite, Fig. \ref{aoft}.
%fffffffffffffffffffffffffffffffffffffffffffffff

\begin{figure}
%\centering
\includegraphics[width=.99\textwidth]{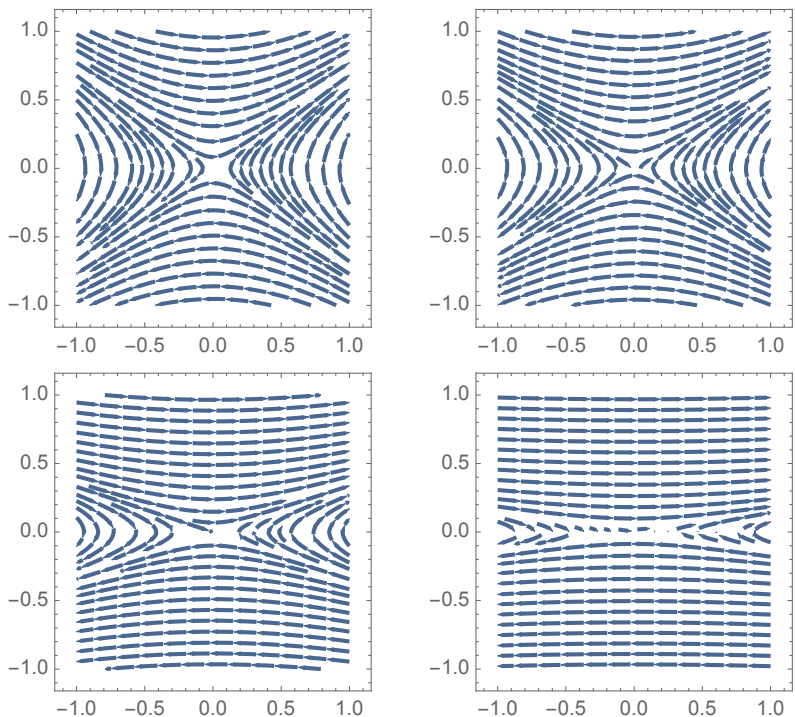}
\caption{Structure of the \Bf\ in the $x-y$ plane during X-point collapse in force-free plasma. The initial configuration on the left is slightly ``squeezed'', $\lambda =0.9$. On dynamical time scale the X-point collapses to form  a current sheet, right Fig. The structure of the \Ef\ in the $x-y$ plane  does not change during the collapse and qualitatively  resembles the $t=0$ configuration of the \Bf.}
\label{B-field-coll} 
\end{figure}

 %  see Fig. \ref{trajectories}.
  (For $\lambda >1$  in finite time $a$ becomes zero, so that $b$ becomes infinite.) Thus, we have generalized the classic solution of \cite{1967JETP...25..656I} to the case of force-free plasma.
 At the moment  when one of the parameters $a$ or $b$ becomes zero, the  current sheet forms,  see Fig. \ref{B-field-coll}. 
 \footnote{In this analysis, the perturbation is of a rather specific type -  a compression on scales well exceeding $L$. We 
probed the reaction of the X-point to small-scale perturbations, of wavelengths $\lambda \ll L$, using numerical 
simulations. The X-point appears to be stable to such perturbations  (see Appendix \ref{un-stressed}). We have also verified that the evolution of the system in case  of  slowly developing  or periodic stress proceeds in a similar fashion - periodic  reversing the stress does not stop the collapse.}

%fffffffffffffffffffffffffffffffffffffffffffffff
%\begin{figure}
%\centering
%\includegraphics[width=.3\textwidth]{aoft.eps}
%\includegraphics[width=.3\textwidth]{trajectories.eps}
%\includegraphics[width=.3\textwidth]{trajectories-1.eps}
%\caption{{\it Left Panel}: An example of the evolution of the parameter $a(t)$, Eq. (\ref{addot}). Initially an X-point is %squeezed by ten percent, $\lambda =0.9$,  parameter ${\cal A}=1$. Evolution occurs on the dynamical time scale, until a %singularity  at  $t=1.42$, so that   the fast growing  stage of the collapse proceeds much quicker. 
%{\it Center Panel.} Particles' drift trajectories during the collapse of the X-point. Initially particles are located on a unit circle.  %(Different fluid elements do not intersect: trajectories show time evolution of each fluid element.) {\it Right Panel} Temporal %evolution of the initially isotropic distribution of particles.  During the X-point collapse, the particles are ``squeezed'' towards %the neural layer.}
%\label{trajectories} 
%\end{figure}
%fffffffffffffffffffffffffffffffffffffffffffffff

For small initial deformation of the X-point, $ \lambda=1-\epsilon$ where $\epsilon \ll 1$ is a small parameter. 
Given the initial conditions   $a(0)=1, \, \dot{a}(0)=0$,  the corresponding asymptotic solution of Eq.\ref{addot} is 
 \be
a \simeq 1 + \epsilon \sinh ^2 t  \,.
\label{eq:x-sol}
 \ee
 This solution is not uniform as at $t_c \approx (1/2)\ln(1/\epsilon)$ the perturbation becomes large. This 
 sets the typical time scale of the X-point collapse.  Since the unit time is $L/c$, where $L$ is the distance where 
 the guide field has the same strength as the in-plane magnetic field of the X-point, it is obvious that with vanishing 
 guide field the collapse occurs instantaneously.   
 
 In this solution, the \Ef\ grows as 
 \be
 \E \simeq \epsilon ( y,x,  - { x^2+y^2 } ) \sinh 2t \,,
 \label{EE}
 \ee
 indicating that at $t\approx t_c$ it may become comparable to the magnetic field. 
Importantly,  the ratio of the \Ef, dominated mostly by $E_z$, to the \Bf, dominated at late times by $B_x$, increases with $y$ (distances away from the newly forming current sheet), 
\be
{E_z \over B_x}  \propto y
\ee
Thus, the analytical model predicts that the \Ef\ becomes of the order of the \Bf\ in a large  domain,  not only  close to the null line.  This is confirmed by  numerical simulations,  see Sections 
\ref{FF-X} and \ref{PIC1}. 
 
 \subsection{Charge starvation during collapse}\label{sec:charge}

During the X-point collapse the \EM\ fields and currents experience a sharp growth, especially near the null line. The current density at the null line  grows exponentially at early times,
 \be
   \bJ \simeq \frac{1}{2\pi} \epsilon \cosh^2 \!t\, {\bf e}_z
 \ee
 Since  $a \rightarrow  \infty$ during collapse, the current density  similarly  increases during the collapse.
 As the parameter $a$ grows, 
 this imposes larger and larger demand on the velocity of the current-carrying particles.  But the maximum current density cannot exceed $ J_z < 2 e  n_e c$.   Thus, if 
 \be
 a(t) >   \sqrt{ L \over \delta} {1\over \sigma^{1/4}}
 \ee
 the current becomes charge-starved (here, $\delta =  c/\om_p, \, \om_p^2 =  4 \pi n_e e^2 /m_e$ are plasma skin depth and plasma frequencies)
 % $\sigma =B_\perp^2 /( 4 \pi n_e m_e c^2)$ is the plasma magnetization parameter 
 %\lorenzo{I usually define sigma with $4\pi$ instead of $8 \pi$, which we should have in order to compare magnetic energy vs plasma energy. if you want, I can change all definitions and numbers}. 
This charge starvation  lead to efficient linear particle acceleration. This is the one of the key points of the model.

 The analytical estimates given above are in agreement with PIC simulations, \S \ref{PIC1}. Our typical runs have $L/ \delta \sim $ hundreds, while the magnetization parameter at the outer scale is 
$\sigma \sim $ thousands.  Thus we expect that the charge starvation occurs approximately at $a \sim $ few -- when the opening of the X-point becomes tens of degrees.

% At the origin, 
 %the displacement current vanishes and the conductivity current 
%$$
  %   \bJ \simeq \frac{1}{2\pi} \epsilon \cosh^2 \!t\, {\bf e}_z
%$$
%becomes large by $t=t_c$. It may exceed the maximum value,  $ J_{max}=2 e  n_e c$, that can be supported by 
%charged particles, leading to the charge starvation regime and hence an efficient linear particle acceleration. 

%% file: FF-X.tex
\section{Force-free  simulations}
\label{FF-X}

The approximate nature of the analytical solution described in the previous section invites a numerical study of the 
X-point collapse in the force-free approximation. To this aim we use a computer code, which solves 
Maxwell equations supplemented with the Ohm law
\begin{equation}
\bJ = \rho\frac{\vpr{E}{B}}{B^2}c + \kappa_\parallel \bE_\parallel
 + \kappa_\perp \bE_\perp.
\label{Ohm1}
\end{equation}
In this equation, the first term represents the drift current (cf. eq.\ref{FF}),  whereas
the second and third terms introduce conductivity along and perpendicular to the magnetic field respectively. 
The parallel conductivity $\kappa_\parallel$ is always set to a high value in order to quickly drive the solution 
towards the force-free state where $E_\parallel =0$. The perpendicular conductivity $\kappa_\perp$ is normally 
set to zero. Only when $E>B$ it is set to a high value in order to quickly obtain $E\le B$.  These two terms also 
introduce dissipation, which becomes significant inside current sheets.  This phenomenological  approach adopted from resistive MHD becomes inaccurate inside collisionless current sheets where plasma effects determine the electric 
current and dissipation. This becomes manifest when we compare our FF and PIC simulations.   The numerical 
scheme is described in details in \citet{2007MNRAS.374..415K}. It is second order in space and time, 
with the source terms treated using the Strand-splitting algorithm. The method of Generalized Lagrange Multiplier (GLM) is employed to keep the magnetic field almost divergence-free. 

The X-point collapse simulations are initialized with 
the magnetic field described by Eq. (\ref{eq:coll}), with parameters $a=1$, $\lambda=\sqrt{0.5}$, and 
vanishing electric field. In the first simulations, we focus on the time-scale corresponding to the onset of 
the X-point collapse, $t <1$.  We use a two-dimensional uniform Cartesian grid with $400\times400$ cells covering the 
x-y domain $[-2,2]\times[-2,2]$ and impose the zero-gradient boundary conditions. Such boundary conditions inevitably  lead to an additional perturbation of the X-point configuration near the boundaries, 
which send waves propagate inside the computational domain with the speed of light. However, these waves do not reach
the central area of interest,  $[-1,1]\times[-1,1]$, on the simulation time scale.  Given the rather strong initial compression 
of the X-point, with $\epsilon \approx 0.4$, the collapse time predicted by the theory is $t_c \approx 0.44$.

%fffffffffffffffffffffffffffffffffffffffffffffffffffffff
\begin{figure}[h!]
\centering
\includegraphics[width=.6\textwidth]{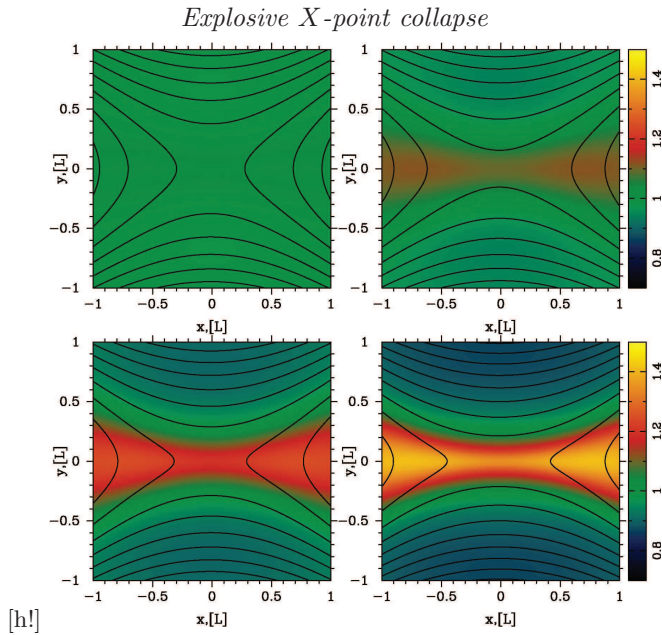}
\caption{Initial phase of a solitary X-point collapse in FF simulations. 
The plots show $B_z$ at $t=$0.25, 0.5, 0.75 and 1. These plots are to be compared 
with  Fig. \ref{x-inner-pic}, which shows the results of PIC simulations with the same
initial setup.
}
\label{x-inner-ff}
\end{figure}
%fffffffffffffffffffffffffffffffffffffffffffffffffffffff  

Fig. \ref{x-inner-ff} shows the magnetic field lines and the strength of the guide field $B_0$ in the central area at four instances during the time interval $[0,1]$.  In accord with with the theory, the degree of the X-point compression increases
with time and becomes visible to naked eye at $t \simeq t_c$. However, the figure also shows that on this time scale the numerical solution begins to deviate strongly from the analytical one. Indeed, the distribution of the guide field becomes significantly non-uniform -- it gets compressed in the plane of collapse ( $y=0$ ).   

\begin{figure}[h!]
\centering
\includegraphics[width=.3\textwidth]{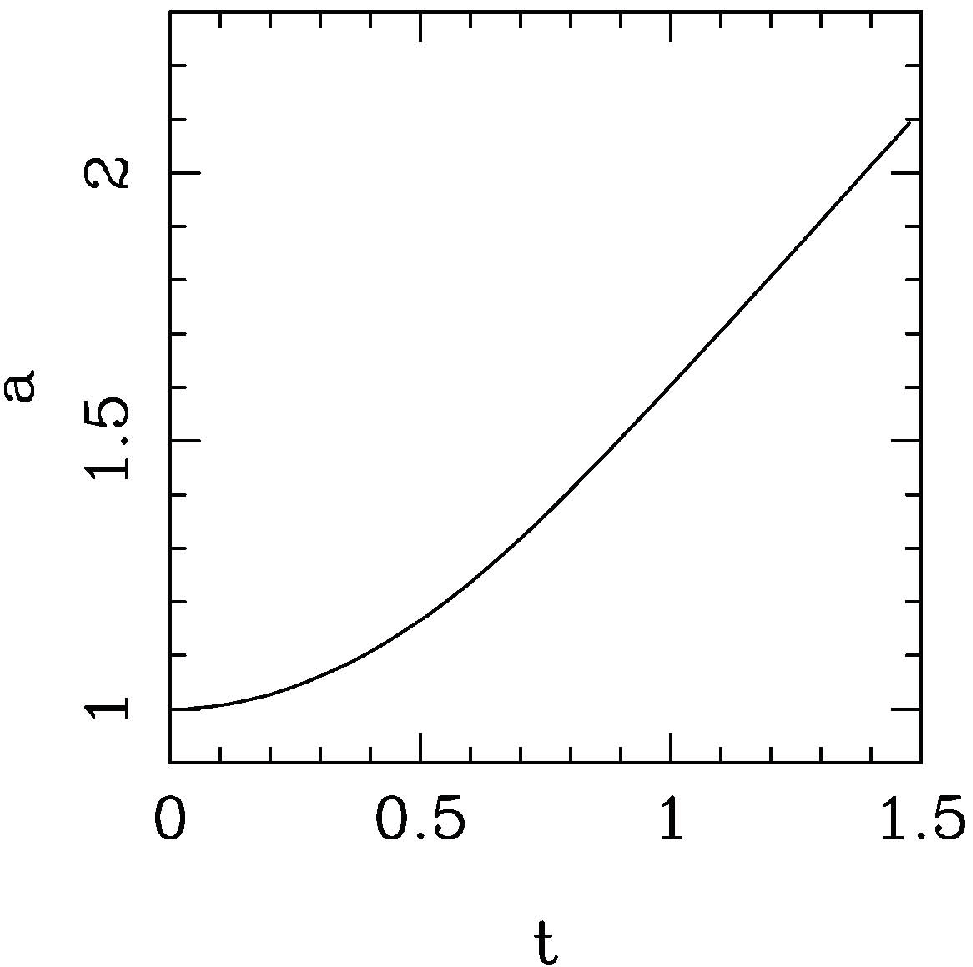}
\includegraphics[width=.3\textwidth]{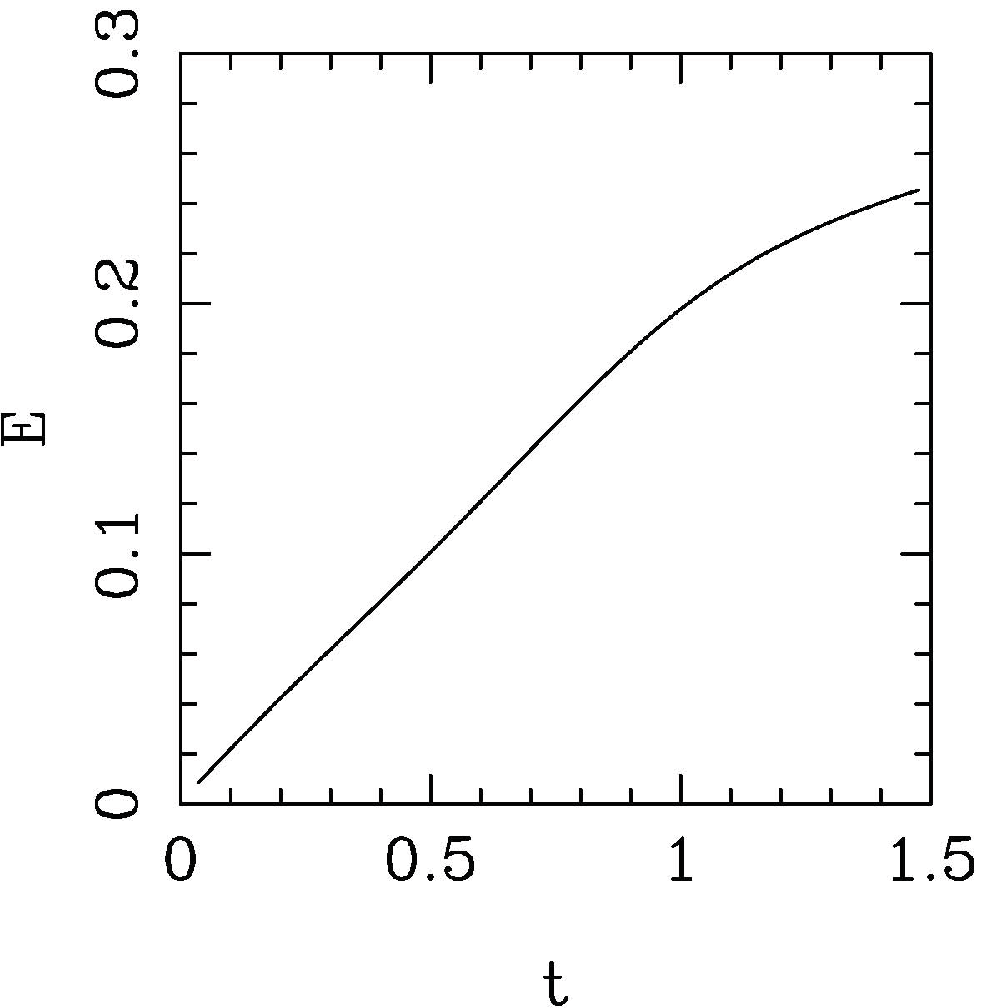}
\caption{Evolution of the parameter $a(t)$ (left panel) and the total electric field strength 
$E(t)$ (right panel) during the initial phase. 
The measurements are taken at the point $(x,y)=(-0.1,0.1)$. 
The analytical solution gives the collapse time $\tau=1.0$. 
These results are sufficiently close, considering the fact that Eq. (\ref{addot}) 
was derived as an asymptotic limit near the $X$-point.}
\label{a-t} 
\end{figure}

This is confirmed in Fig. \ref{a-t}, which shows the evolution of the compression parameter $a$ and the 
electric field strength,  both computed at the point $(x,y)=(-0.1,0.1)$. In order to measure the local value of $a$,
we use use Eq.\ref{eq:coll}, which yields  
$$
a(t)=\lambda^{1/2}(xB_x/yB_y)^{1/4}\,. 
$$ 
One can see that although the characteristic time scale is close to $t_c$,
the numerical solution does not quite follow the analytical one. This is 
expected as the analytic solution is only accurate for $t\ll t_c$.  The PIC simulations, which are described in 
the next section, yield very similar results.    

%fffffffffffffffffffffffffffffffffffffffffffffffffffffff
%fffffffffffffffffffffffffffffffffffffffffffffffffffffff

%fffffffffffffffffffffffffffffffffffffffffffffffffffffff
\begin{figure}[h!]
\centering
\includegraphics[width=.49\textwidth]{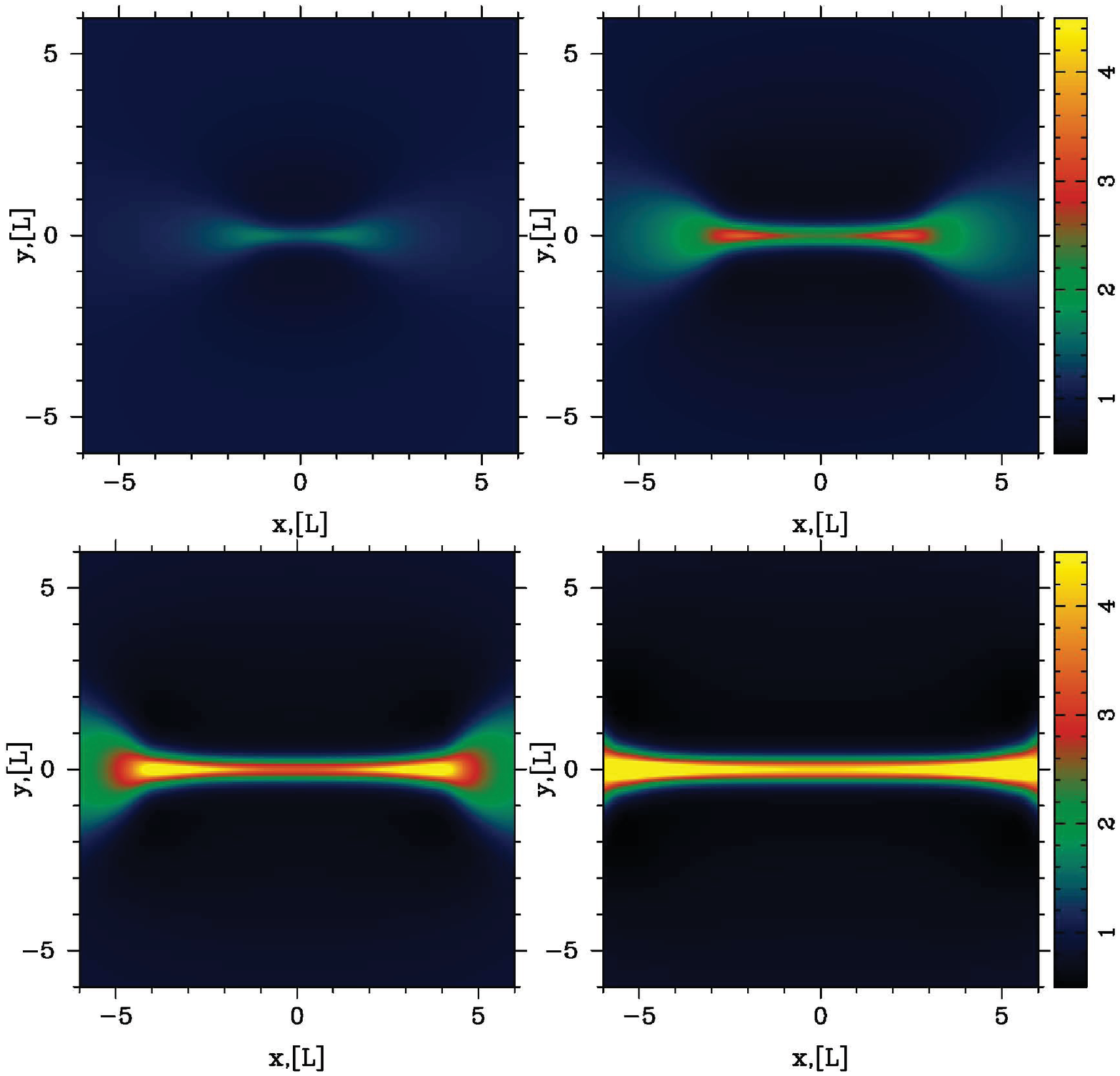}
\includegraphics[width=.49\textwidth]{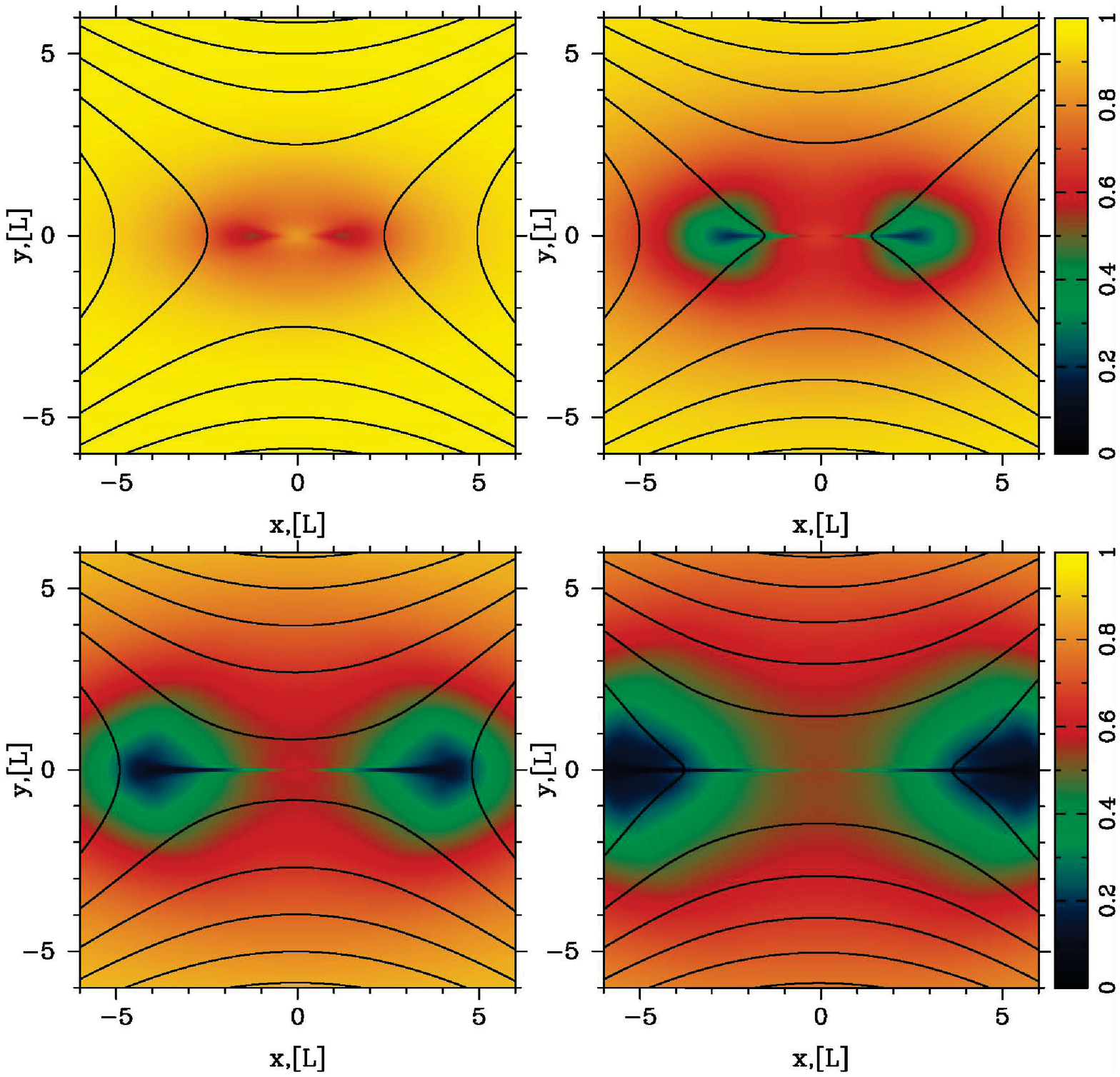}
\caption{Long-term evolution of stressed  solitary X-point in FF simulations. 
The left panel shows the $B_z$ component of the magnetic field. 
The right panel shows $1-E^2/B^2$ (colour) and magnetic field lines
% \lorenzo{just to be sure: in my plots I include all components of E and B, are you doing the same?}. 
The plots show the numerical solution at $t=1.5,$ 3, 4.5 and 6.
PIC simulations for the same initial configuration are shown in 
Figures~\ref{pic-xcg-b3} and \ref{pic-xcg-bme}.
}
\label{x-outer-ff} 
\end{figure}
%fffffffffffffffffffffffffffffffffffffffffffffffffffffff
In order to study the evolution of the X-point at $t>t_c$, we repeated the simulations on 
in a larger computational domain, $[-10,10]\times[-10,10]$ with $800\times800$ cells.
The results are illustrated in Fig. \ref{x-outer-ff}. One can see that at $t\approx t_c$ the X-point turns 
into a current sheet, bounded by two Y-points.  The separation between these Y-points increase with approximately 
twice the speed of light. For  
$t\gg t_c$ the solution begins to exhibits a self-similar evolution, which is expected as 
the only characteristic length scale of this problem is $l=1$. 
Ahead of each of Y-point there are bow-shaped regions where the drift speed is very 
close to $c$ and  $E\simeq B$.  Inside the current sheet the force-free approximation breaks down 
completely as the electric field strengh tends to exceed that of the magnetic field. Given our 
prescription for the resistivity some of the electromagnetic field  disappears inside the current sheet 
without a trace. In order to capture the evolution of this current sheet properly particles must be 
reintroduced into the system, which done in the PIC simulations described in the next section.

%% file: PIC1.tex
 
\section{PIC simulations}
\label{PIC1}

\subsection{Overall principles and parameters of PIC simulations}

The most fundamental method for simulating the kinetic dynamics of a
reconnecting plasma involves the use of particle-in-cell (PIC)
codes, that evolve the discretized equations of electrodynamics ---
Maxwell's equations and the Lorentz force law \citep{hockney,birdsall}. PIC
codes can model astrophysical plasmas from first principles, as a
collection of charged macro-particles --- each representing many physical
particles --- that are moved by integration
of the Lorentz force.\footnote{For this work, we employ the Vay pusher, since we find that it is more accurate than the standard Boris algorithm in dealing with the relativistic drift velocities associated with the reconnection flows \citep{vay_08}.}
 The electric currents associated with the macro-particles are
deposited on a grid, where Maxwell's equations are
discretized. Electromagnetic fields are then advanced via Maxwell's
equations, with particle currents as the source term. Finally, the
updated fields are extrapolated to the particle locations and used
for the computation of the Lorentz force, so the loop is closed
self-consistently.

We study the collapse of a solitary X-point with 2D PIC simulations, employing the massively parallel electromagnetic PIC code  TRISTAN-MP \citep{buneman_93,spitkovsky_05}. 
Our computational domain is a square in the $x-y$ plane, which is initialized with a uniform density of cold electron-positron plasma (the composition of pulsar wind nebulae is inferred to have negligible ion/proton component). The simulation is initialized with a vanishing electric field and with the magnetic field appropriate for a stressed X-point collapse (see Eq. (\ref{eq:coll})) with $\lambda=1/\sqrt{2}$, for direct comparison with the force-free results described above.\footnote{We have also explicitly verified that an unstressed X-point (i.e., with $\lambda=1$) is stable, Appendix  \ref{un-stressed}.} The stressed X-point configuration would require a particle current in the direction perpendicular to the simulation plane, to sustain the curl of the field (which, on the other hand, is absent in the case of an unstressed X-point). In our setup, the computational particles are initialized at rest, but such electric current gets self-consistently built up within a few timesteps. At the boundaries of the box, the field is reset at every timestep to the initial configuration. This leads to an artificial deformation of the self-consistent X-point evolution which propagates from the boundaries toward the center at the speed of light. However, our domain is chosen to be large enough such that this perturbation does not reach the central area of interest within the timespan covered by our simulations.

We perform two sets of simulations.  First, we explore the physics at relatively early times with a 2D Cartesian grid of $32768\times 32768$ cells. The spatial resolution is such that the plasma skin depth $\comp$ is resolved with 10 cells.\footnote{In the case of a cold plasma, the  skin depth is defined as $\comp=\sqrt{mc^2/4 \pi n e^2}$. For a hot plasma, it is defined as $\comp=\sqrt{mc^2[1+(\hat{\gamma}-1)^{-1} kT/m c^2]/4 \pi n e^2}$, where $\hat{\gamma}$ is the adiabatic index.} The unit length is $L=800\,c/\omega_{\rm p}$, so that the domain size is a square with $\simeq 4L$ on each side. The physical region of interest is the central $2L\times 2L$ square. The simulation is evolved up to $\sim L/c$, so that the artificial perturbation driven by the boundaries does not affect the region of interest. In a second set of simulations, we explore the physics at late times, with a 2D Cartesian box of $40960\times 40960$ cells, with spatial resolution $\comp=1.25$ cells. With the unit length still being $L=800\,c/\omega_{\rm p}$, the overall system is a square of $\simeq 41 L$ on each side. At early times, the evolution is entirely consistent with the results of the first set of simulations described above, which suggests that a spatial resolution of $\comp=1.25$ cells is still sufficient to capture the relevant physics (more generally, we have checked for numerical convergence from $\comp=1.25$ cells up to $\comp=20$ cells). We follow the large-scale system up to $\sim 6 L/c$, so that the evolution of the physical region at the center stays unaffected by the boundary conditions. 

For both sets of simulations, each cell is initialized with two macro-positrons and two macro-electrons (with a small thermal dispersion of $kT/m c^2=10^{-4}$). Hence the initial plasma density distribution is uniform, whereas the magnetic field is not. This leads to a non-uniform magnetization, which increases with the distance from the centre line of the X-point. 
We have checked that our results are the same when using up to 24 particles per cell.  In order to reduce noise in the simulation, we filter the electric current deposited onto the grid by the particles, effectively mimicking the effect of a larger number of particles per cell \citep{spitkovsky_05,belyaev_15}. 

We explore the dependence of our results on two physical parameters, namely the strength of the guide field and the flow magnetization. In the simulations with guide field, the guide field is initially uniform and its strength is chosen to be equal to that of the unstressed in-plane field of the X-point at the unit distance from the origin.  This case allows for a direct comparison with analytical theory and force-free results. We also studied the case without a guide field. This case will be relevant for our investigation of the collapse of ABC structures, considered in the second paper of this series. There, we will demonstrate that particle acceleration is most efficient at null points, i.e., where both the in-plane fields and the out-of-plane guide field vanish. 

In addition to the guide field strength, we investigate the dependence of our results on the flow magnetization, which for a cold electron-positron  plasma reduces to $\sigma=B^2/4\pi m n c^2$. The physics of X-point collapse has been
widely studied in the literature with PIC simulations \citep{tsiklauri_4,tsiklauri_3,tsiklauri_2,tsiklauri_1}, but only in the non-relativistic regime $\sigma\ll1$. To the best of our knowledge, our investigation is the first to focus on the relativistic regime $\sigma\gg1$, which is appropriate for relativistic astrophysical outflows. 
Since the in-plane field of the initial configuration grows linearly with distance from the centre line and the plasma density is uniform, the corresponding magnetization increases quadratically with the distance in the case without guide field 
and somewhat slower when the guide field is included.  For definiteness,  we opted to parametrise our runs via the initial value $\sigma_L$  of plasma magnetization  at the unit distance $L$ from the origin (along the unstressed $x$ direction), measured only with the in-plane fields (so, excluding the guide field). We explore three values of $\sigma_L$: $4\times10^2$, $4\times 10^3$ and $4\times 10^4$.

%%%%%%%%%%%%%%%%%%%%%%%
\begin{figure}[!ht]
\centering
\includegraphics[width=.49\textwidth]{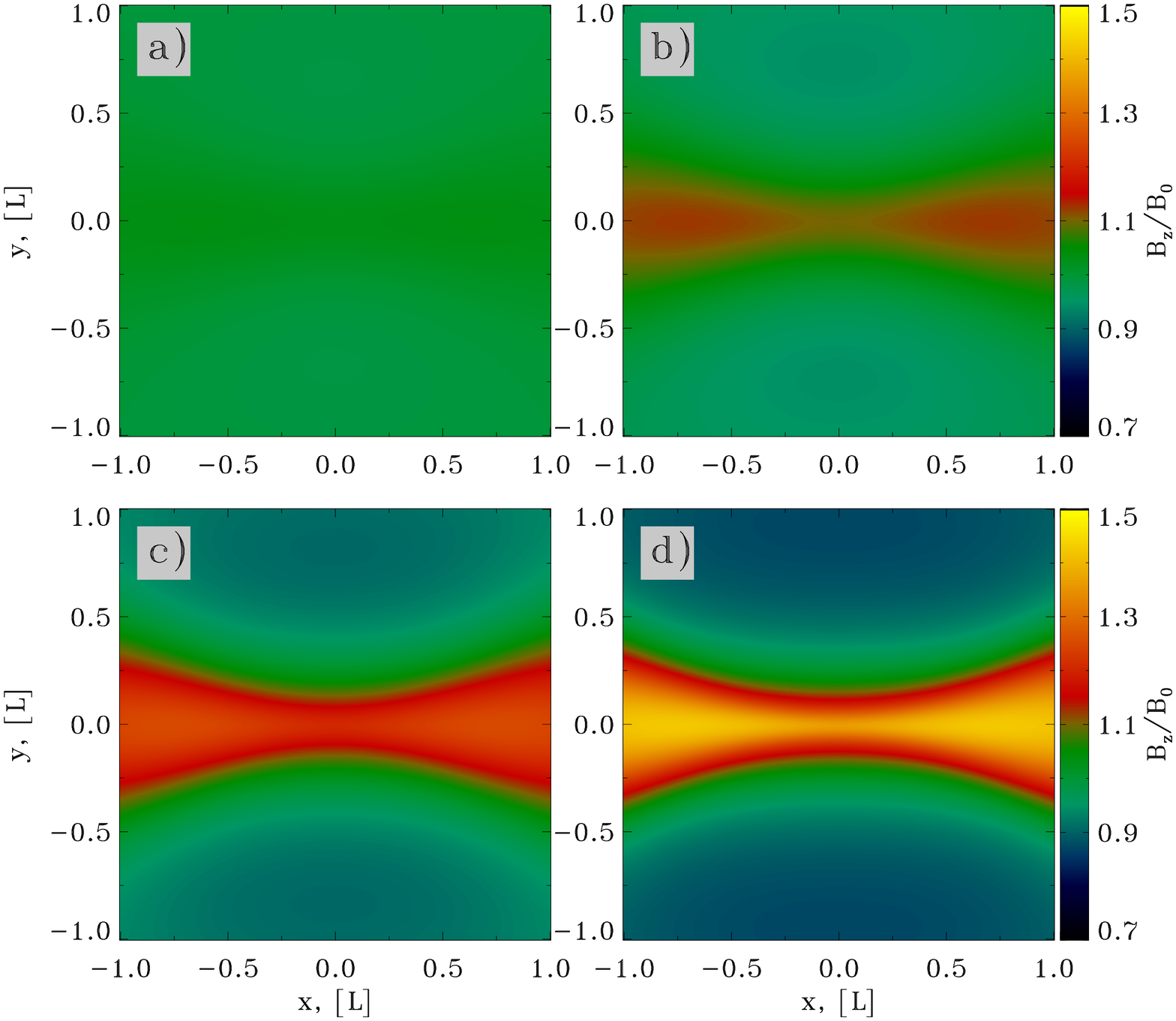} 
\includegraphics[width=.49\textwidth]{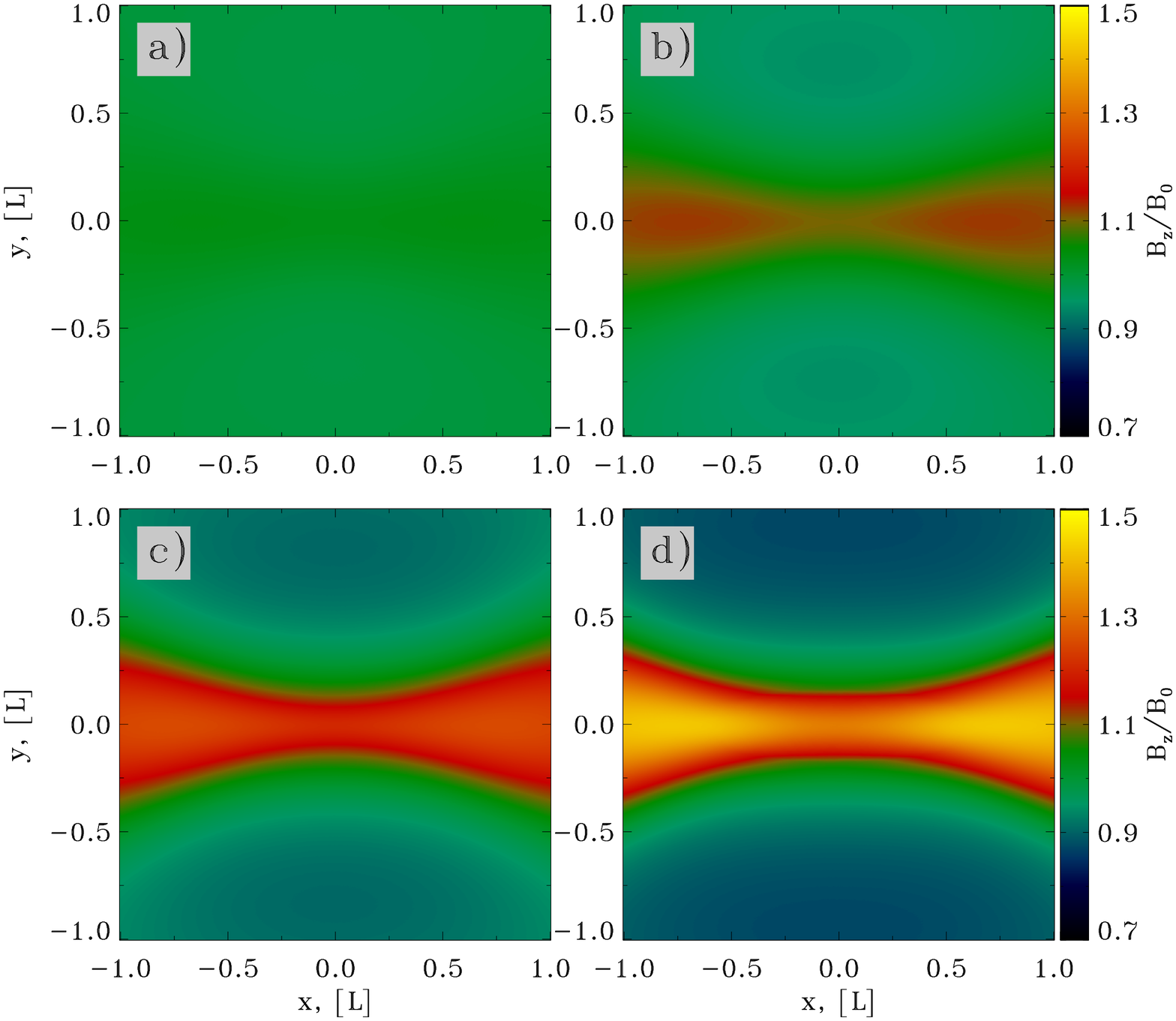} 
\caption{Initial phase of an X-point collapse in PIC simulations with guide field, for two different 
magnetizations: $\sigma_L =4\ex{3}$  (left)  and  $\sigma_L =4\ex{4}$  (right). 
The plots show the out-of-plane field $B_z$ at $ct/L=$0.25, 0.5, 0.75 and 1, from panel (a) to (d). This figure corresponds to Fig.~\ref{x-inner-ff}, which shows the results of force-free simulations.}
\label{x-inner-pic} 
\end{figure}

 \begin{figure}[!ht]
 \centering
\includegraphics[width=.49\textwidth]{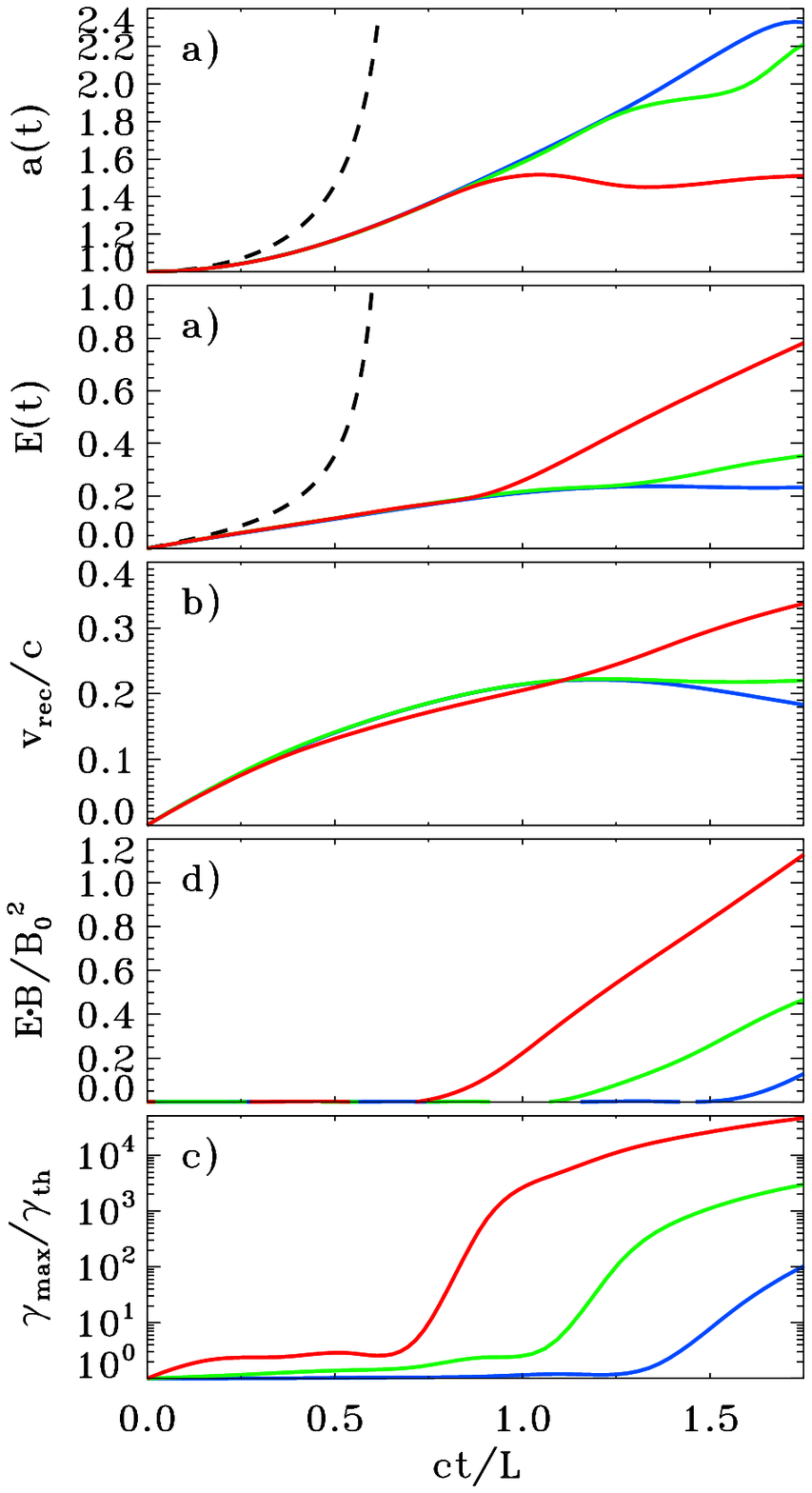} 
\caption{Temporal evolution of various quantities from PIC simulations of an X-point collapse with guide field, for three values of the magnetization: $\sigma_L =4\times10^2$  (blue), $\sigma_L =4\ex{3}$  (green)  and  $\sigma_L =4\ex{4}$ (red). As a function of time, we plot: (a) the value of $a(t)=\lambda^{1/2}(B_x/B_y)^{1/4}$ at the location $(-0.1L,0.1L)$, to be compared with the result  of force-free simulations in the left panel of Fig.~\ref{a-t} and with the analytical estimates (dashed line);  (b) the value of the electric field strength $E(t)$ at the location $(-0.1L,0.1L)$ in units of $B_0$, to be compared with the result  of force-free simulations in the right panel of Fig.~\ref{a-t} and with the analytical estimates (dashed line); (c), the reconnection rate, defined as the  inflow speed along the $y$ direction averaged over a square of side equal to $L$ around the central region; (d) the parameter $\bmath{E}\cdot\bmath{B}/B_0^2$ at the center of the domain, which explicitly shows when the force-free condition $\bmath{E}\cdot\bmath{B}=0$ is broken; (d) the maximum particle Lorentz factor $\gamma_{\rm max}$ (as defined in \eq{ggmax}), in units of the thermal value $\gamma_{\rm th}\simeq 1+(\hat{\gamma}-1)^{-1} kT/m c^2$, which in this case of a cold plasma reduces to $\gamma_{\rm th}\simeq 1$.}
\label{fig:xguidetimecomp} 
\end{figure}

 \begin{figure}[!ht]
 \centering
\includegraphics[width=.49\textwidth]{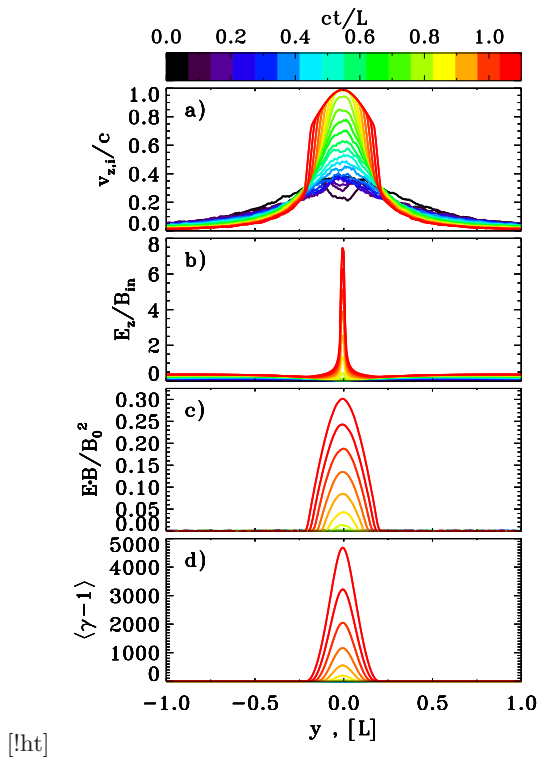} 
\caption{Spatial profiles of various quantities from a PIC simulation of an X-point collapse with guide field and magnetization $\sigma_L=4\times 10^4$, which corresponds to the red curves in \fig{xguidetimecomp}. As a function of the coordinate $y$ along the inflow direction, we plot at $x=0$: (a) the bulk speed of positrons, in units of the speed of light (the bulk speed of electrons is equal and opposite); (b) the ratio of the out-of-plane electric field $E_z$ to the in-plane magnetic field $B_{in}=\sqrt{B_x^2+B_y^2}$; (c) the parameter $\bmath{E}\cdot\bmath{B}/B_0^2$, which explicitly shows when the force-free condition $\bmath{E}\cdot\bmath{B}=0$ is falsified; (d) the mean particle Lorentz factor.}
\label{fig:xguidespace} 
\end{figure}

%%%%%%%%%%%%%%%%%%%%%%%
 \begin{figure}[!ht]
 \centering
\includegraphics[width=.49\textwidth]{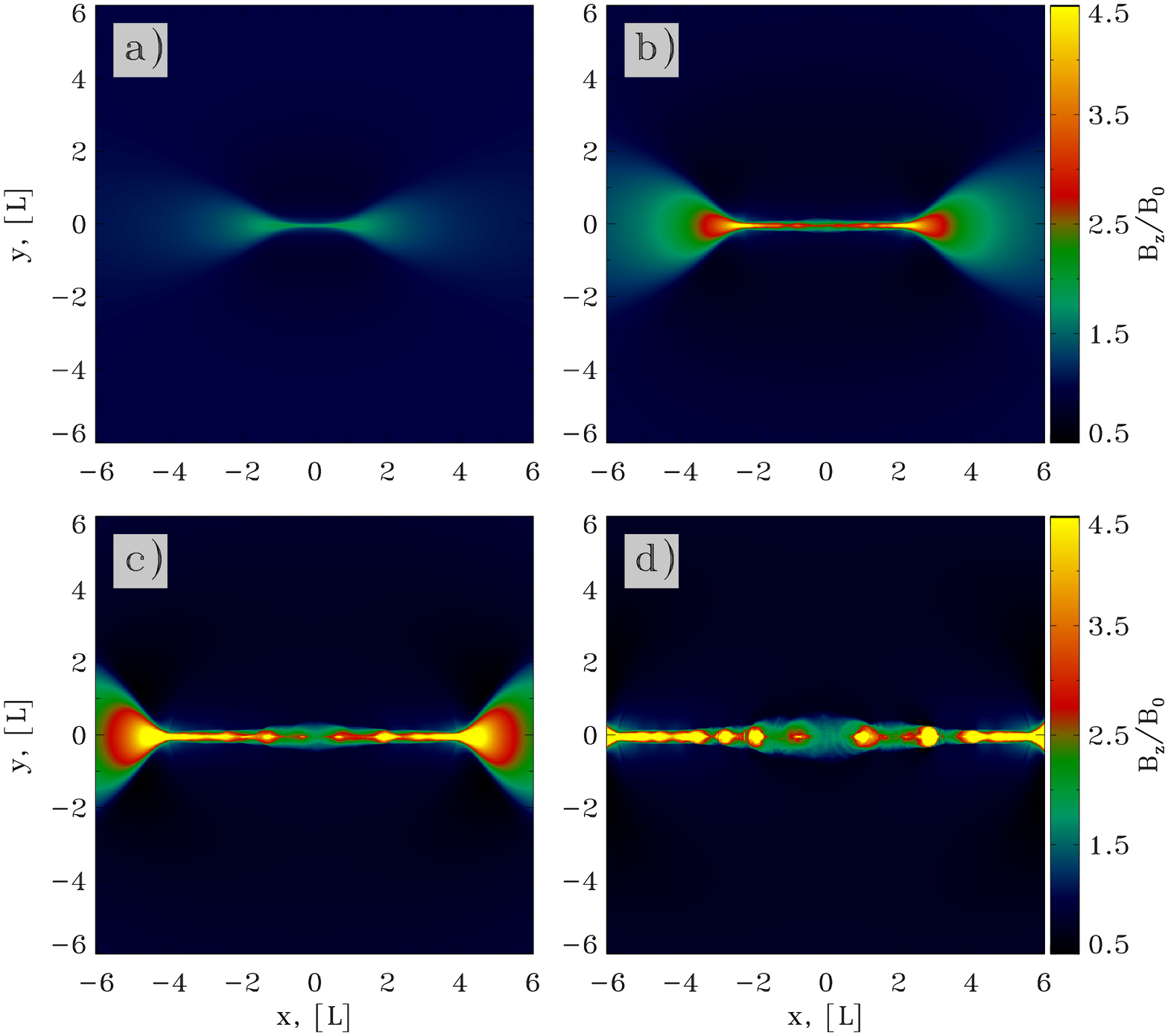} 
\includegraphics[width=.49\textwidth]{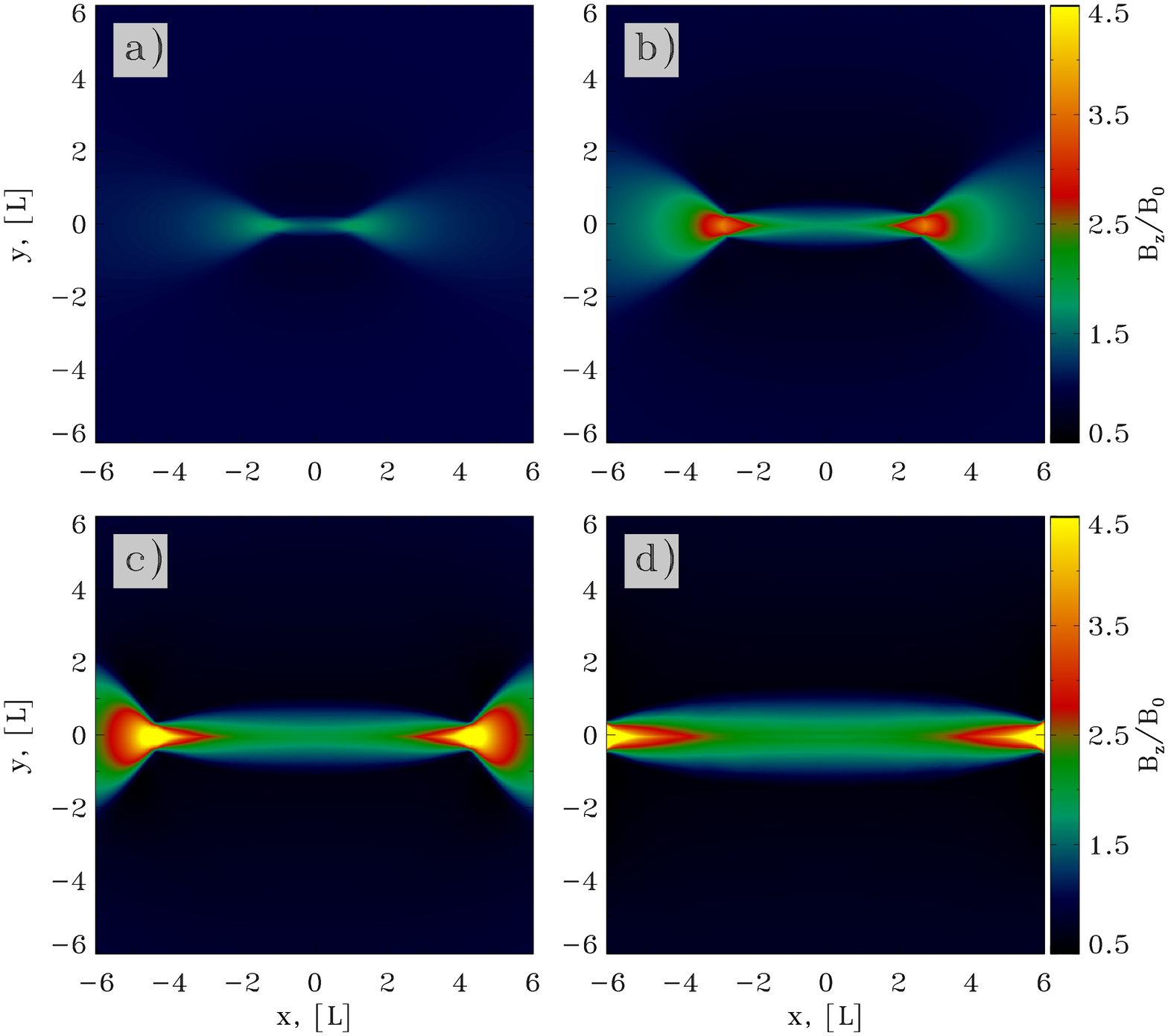} 
\caption{Late time evolution of the X-point collapse in PIC simulations with guide field, for two different magnetizations: $\sigma_L =4\ex{3}$  (left)  and  $\sigma_L =4\ex{4}$  (right). The plots show the out-of-plane field at $ct/L=$1.5, 3, 4.5, 6, from panel (a) to (d). This figure corresponds to the left side of Fig.~\ref{x-outer-ff}, which shows the results of force-free simulations.}
\label{pic-xcg-b3} 
\end{figure}

 \begin{figure}[!ht]
 \centering
\includegraphics[width=.49\textwidth]{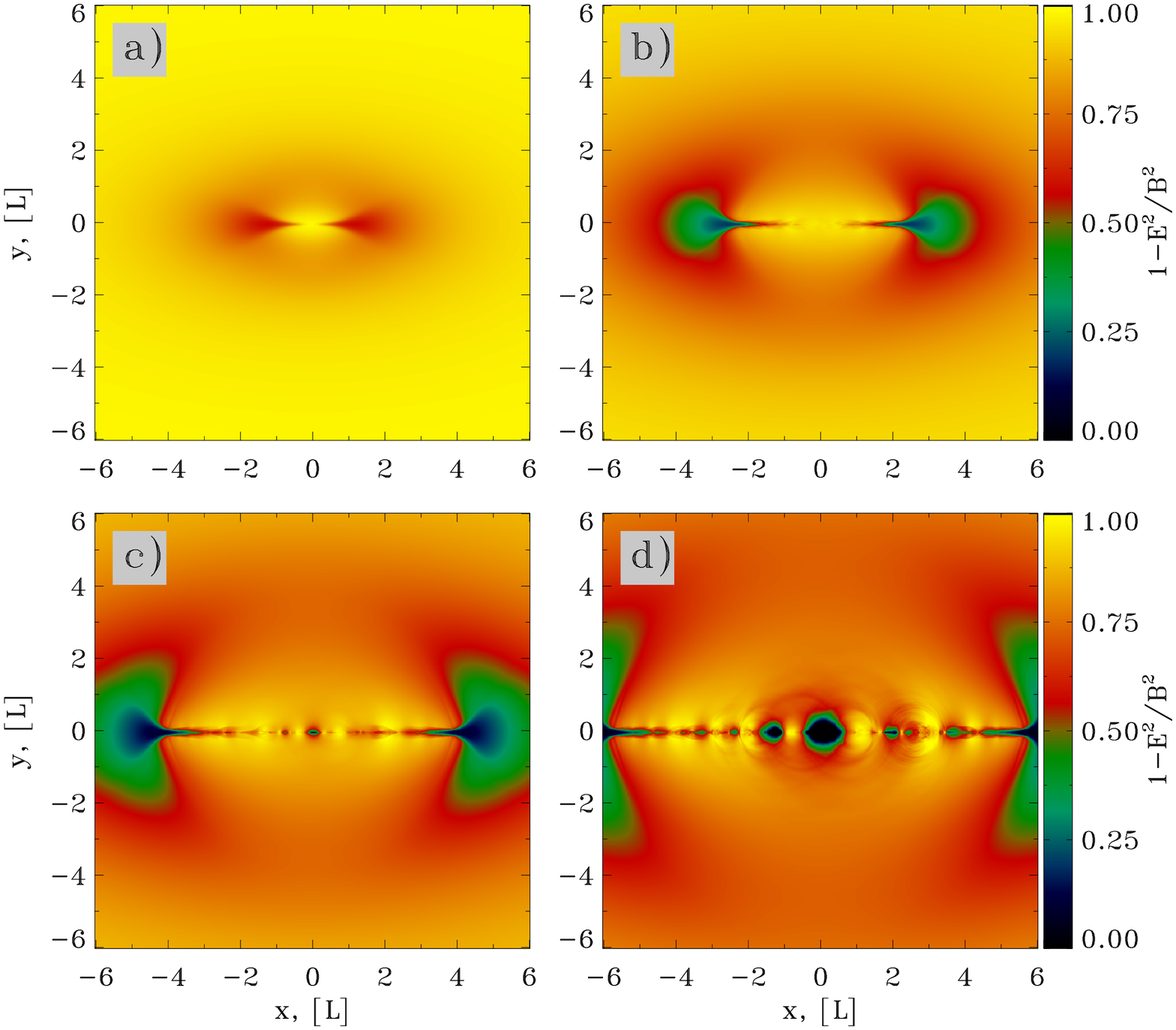} 
\includegraphics[width=.49\textwidth]{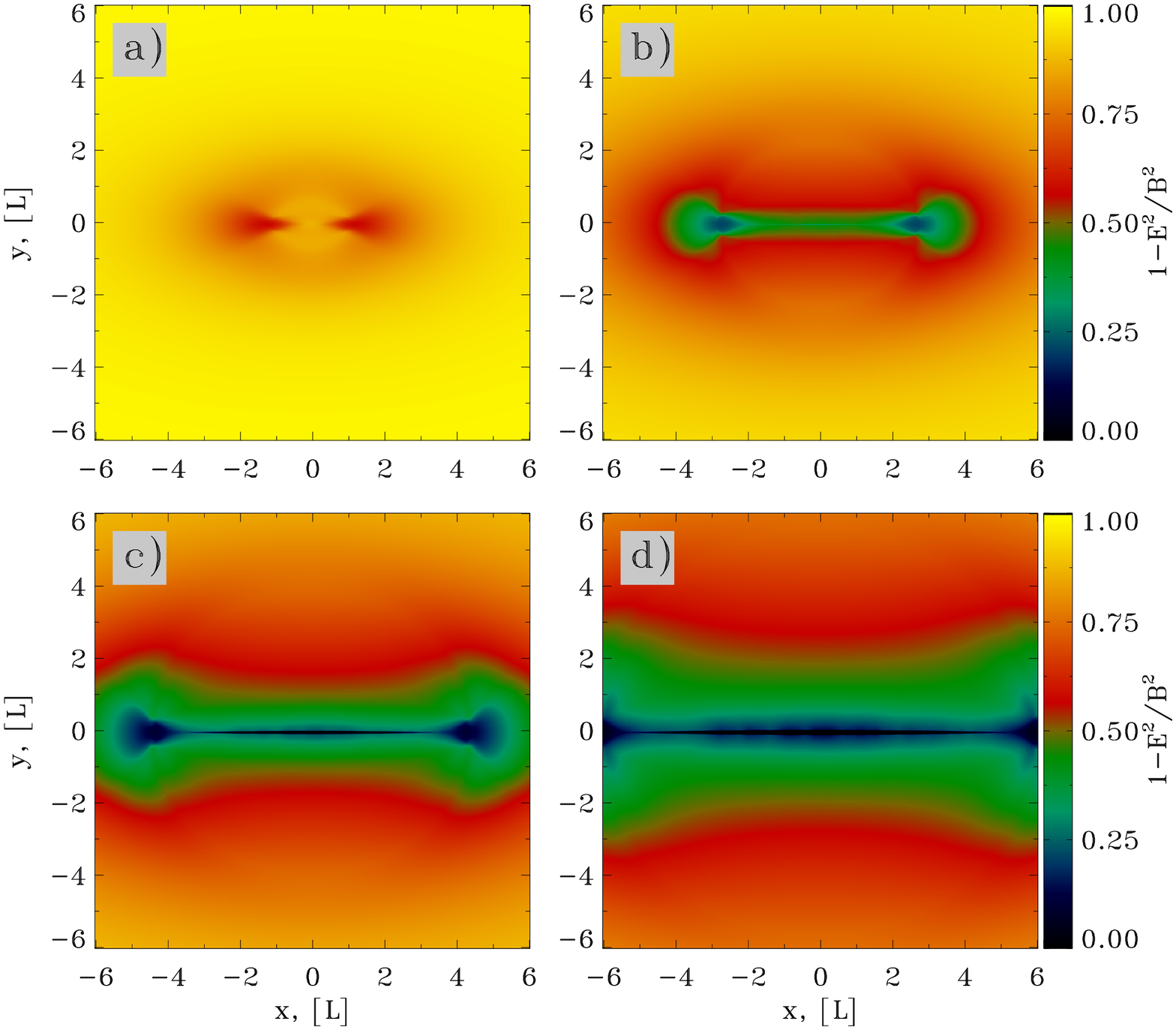} 
\caption{Late time evolution of the X-point collapse in PIC simulations with guide field, for two different magnetizations: $\sigma_L =4\ex{3}$  (left)  and  $\sigma_L =4\ex{4}$  (right). The plots show the quantity $1-E^2/B^2$ at $ct/L=$1.5, 3, 4.5, 6, from panel (a) to (d) (strictly speaking, we plot $\max[0,1-E^2/B^2]$, for direct comparison with force-free simulations, that implicitly constrain $E\leq B$). This figure corresponds to the right side of Fig.~\ref{x-outer-ff}, which shows the results of force-free simulations.}
\label{pic-xcg-bme} 
\end{figure}

%%%%%%%%%%%%%%%%%%%%%%%
\subsection{Stressed X-point collapse with guide field}\label{sec:xguide}

Figure~\ref{x-inner-pic} shows the initial phase  ($ct/L\leq 1$) of the collapse of a solitary X-point in PIC simulations with $\lambda=1/\sqrt{2}$ and with guide field,  for two different 
magnetizations: $\sigma_L =4\ex{3}$  (left)  and  $\sigma_L =4\ex{4}$  (right). The expected rapid collapse of the squeezed X-point is clearly visible, and in excellent agreement with the 2D results of force-free simulations, as shown in Fig.~\ref{x-inner-ff}. The out-of-plane magnetic field $B_z$ is compressed toward the $y=0$ plane, in agreement with the force-free results, but in apparent contradiction with the analytical solution, that assumed no evolution of the guide field. The 2D pattern of $B_z$ in PIC simulations is remarkably independent of the magnetization (compare left and right), for the early phases presented in Fig.~\ref{x-inner-pic}.

The fact that PIC results at early times are independent of $\sigma_L$ is further illustrated in \fig{xguidetimecomp}, where we present the temporal evolution of various quantities  for three values of the magnetization: $\sigma_L =4\times10^2$  (blue), $\sigma_L =4\ex{3}$  (green)  and  $\sigma_L =4\ex{4}$ (red). Both the value of $a(t)=\lambda^{1/2}(B_x/B_y)^{1/4}$ and of the electric field $E(t)$ (in units of $B_0$) at the location $(-0.1L,0.1L)$ are independent of $\sigma_L$, as long as $ct/L\lesssim1$, and they are in excellent agreement with the results of force-free simulations presented in Fig.~\ref{a-t}. In other words, the physics at early times is entirely regulated by large-scale electromagnetic stresses, with no appreciable particle kinetic effects. Carried along by the converging magnetic fields of the collapsing X-point, particles are flowing into the central region, with a reconnection speed of $\sim 0.2c$ (averaged over a square of side equal to $L$ around the center). This is significantly higher compared to the typical reconnection rate for a plane current sheet configuration. For a relativistic current sheet with guide field of the same strength as the alternating field, the reconnection rate is only  $v_{\rm rec}/c\sim 0.02$ (Sironi \& Spitkovsky 2017, in prep.).\footnote{ \citet{liu_15} report a significantly higher reconnection rate. However, their results are consistent with the findings in Sironi \& Spitkovsky 2017, in prep.). The apparent  disagreement is due to the fact that  \citet{liu_15} measured the reconnection rate on the microscopic skin-depth scales, whereas 
Sironi \& Spitkovsky (2017)  on macroscopic scales. Similarly, in the present paper the scale of measurement $L\gg\comp$
is macroscopic.}

We have argued in Sec.\ref{X-point} that as the system evolves and the $a(t)$ parameter increases, the electric current may become charge-starved. In \fig{xguidetimecomp}, this is clearly indicated by the time when the force-free condition $\bmath{E}\cdot\bmath{B}=0$ is violated, as shown in panel (d). Higher values of $\sigma_L$ lead to an earlier onset of charge starvation: the simulation with $\sigma_L=4\times 10^4$ becomes charge starved at $ct/L\simeq 0.75$, the simulation with $\sigma_L=4\times 10^3$ at $ct/L\simeq 1.1$ and the simulation with $\sigma_L=4\times 10^2$ at $ct/L\simeq 1.5$. 
This is expected as higher $\sigma$ corresponds to fewer available charges.
The onset of charge starvation is accompanied by a deviation of the curves in panels (a) and (b) from the results of force-free simulations, where the condition $\bmath{E}\cdot\bmath{B}=0$ is imposed by hand at all times. 

The physics of charge starvation is illustrated in \fig{xguidespace}, for the case $\sigma_L=4\times10^4$ that corresponds to the red curves in \fig{xguidetimecomp}. In response to the rapidly increasing curl of the magnetic field, the $z$ velocity of the charge carriers has to increase (\fig{xguidespace}(a), for positrons), while their density stays nearly unchanged. When the drifting speed reaches the speed of light, at $ct/L\simeq 0.8$ in \fig{xguidespace}(a), the particle electric current cannot sustain the curl of the magnetic field any longer and the displacement current takes over. As a result, the electric field grows to violate the force-free condition $\bmath{E}\cdot\bmath{B}=0$. In fact, panel (c) shows that the value of $\bmath{E}\cdot\bmath{B}$ departs from zero at $ct/L\simeq 0.8$, i.e., right when the particle drift velocity approaches the speed of light. The nonzero $\bmath{E}\cdot\bmath{B}$ triggers the onset of efficient particle acceleration, as shown by the profile of the mean particle Lorentz factor in panel (d). Indeed, the locations of efficient particle acceleration (i.e., where $\langle\gamma\rangle\gg1$) are spatially coincident with the regions where $\bmath{E}\cdot\bmath{B}\neq 0$. There, the pressure of accelerated particles provides a significant back-reaction onto the field collapse, and the agreement with the force-free results necessarily fails.

After the onset of charge starvation, the maximum particle energy dramatically increases (see \fig{xguidetimecomp}(e)). It is this period of rapid acceleration that will be extensively studied in the following sections. Here, and hereafter, the maximum particle Lorentz factor plotted in \fig{xguidetimecomp}(e) is defined as
\be\label{eq:ggmax}
\gamma_{\rm max}=\frac{\int \gamma^{n_{\rm cut}} dN/d\gamma\, d\gamma}{\int \gamma^{n_{\rm cut}-1} dN/d\gamma\, d\gamma}
\ee
where $n_{\rm cut}$ is empirically chosen to be $n_{\rm cut}=6$ \citep[see also][]{bai_15}. If the particle energy spectrum takes the form $dN/d\gamma\propto \gamma^{-s} \exp(-\gamma/\gamma_{\rm cut})$ with power-law slope $s$ and exponential cutoff at $\gamma_{\rm cut}$, then our definition yields $\gammamax\sim (n_{\rm cut}-s)\,\gamma_{\rm cut}$.

As the collapse proceeds beyond $ct/L\sim 1$, the system approaches a self-similar evolution, as we have already emphasized for our force-free simulations (see Fig.~\ref{x-outer-ff}). As shown in Fig.~\ref{pic-xcg-b3}, the X-point evolves into a thin current sheet with two Y-points at its ends, which move with speed very close to the speed of light. The current sheet is supported by the pressure of the compressed guide field (as it is apparent in Fig.~\ref{pic-xcg-b3}) and by the kinetic pressure of the accelerated particles. As illustrated in Fig.~\ref{pic-xcg-b3}, the current sheet is thinner for lower magnetizations, at fixed $L/\comp$ (compare $\sigma_L =4\ex{3}$  on the left  and  $\sigma_L =4\ex{4}$  on the right). Roughly, the thickness of the current sheet at this stage is set by the Larmor radius $r_{\rm L,hot}=\sqrt{\sigma_L}\comp$ of the high-energy particles heated/accelerated by reconnection, which explains the variation of the current sheet thickness with $\sigma_L$ in Fig.~\ref{pic-xcg-b3}. A long thin current sheet is expected to fragment into a chain of plasmoids/magnetic islands \citep[e.g.,][]{uzdensky_10}, when the length-to-thickness ratio is much larger than unity. At fixed time and hence similar sheet length, it is then more likely that the fragmentation into plasmoids appears at lower magnetizations, since a lower $\sigma_L$ results in a thinner current sheet. This is in agreement with Fig.~\ref{pic-xcg-b3}, and we have further checked that the current sheet in the  simulation with $\sigma_L=4\times10^2$ starts fragmenting at even earlier times. 

In the small-scale X-points in between the self-generated plasmoids, the electric field can approach and exceed the 
magnetic field. This is apparent in Fig.~\ref{pic-xcg-bme} --- referring to the same simulations as in Fig.~\ref{pic-xcg-b3} --- where we show the value of $1-E^2/B^2$, which quantifies the strength of the electric field relative to the magnetic field. In the case of $\sigma_L=4\times 10^3$ (left side), the {\it microscopic} regions in between the plasmoids are characterized by $E>B$ (see, e.g., at the center of panel (d)). In addition, ahead of each of the two Y-points, a bow-shaped area exists where $E\sim B$ (e.g., at $|x|\sim 5L$ and $y\sim 0$ in panel (c)). The two Y-points move at the Alfv\'en
speed, which is comparable to the speed of light for our $\sigma_L\gg1$ plasma. So, the fact that $E\sim B$ ahead of the Y-points is just a manifestation of the relativistic
nature of the reconnection outflows. For $\sigma_L=4\ex{3}$ (left side in Fig.~\ref{pic-xcg-bme}),
the electric energy in the bulk of the inflow region is much smaller than the magnetic energy,
corresponding to $1-E^2/B^2\sim 0.6$. Or equivalently, the reconnection speed is significantly smaller than the speed of light.
%\footnote{For such a large value of the magnetization, this might appear surprising. However, in the initial setup, the magnetization close to the X-point is much smaller (in fact, the in-plane field increases linearly with distance). At late times, the pressure of the guide field accumulated in the reconnection layer, together with the particle pressure, prevents the reconnection inflow to achieve relativistic speeds.}  
The highly relativistic case of $\sigma_L=4\ex{4}$ (right panel in
Fig. \ref{pic-xcg-bme}) shows a different picture. Here, a large volume with $E \sim B$ develops in the inflow region. In other words, the reconnection speed approaches the speed of light in a {\it macroscopic} region. In the next subsection, we will show that an inflow speed near the speed of light (or equivalently, $E\sim B$) is a generic by-product of high-$\sigma_L$ reconnection.\footnote{In retrospect, the fact that this conclusion also holds for the case of guide-field X-point collapse is not surprising. At the initial time, only the region within a radius $\lesssim L$ from the current sheet has a guide field stronger than the in-plane fields. This implies that at late times, when regions initially at a distance $\gtrsim L$ are eventually advected to the center, the guide field at the current sheet will be sub-dominant with respect to the in-plane fields, so that the results of guide-field reconnection will resemble the case of a vanishing guide field.}

%%%%%%%%%%%%%%%%%%%%%%%
 \begin{figure}[!ht]
 \centering
\includegraphics[width=.49\textwidth]{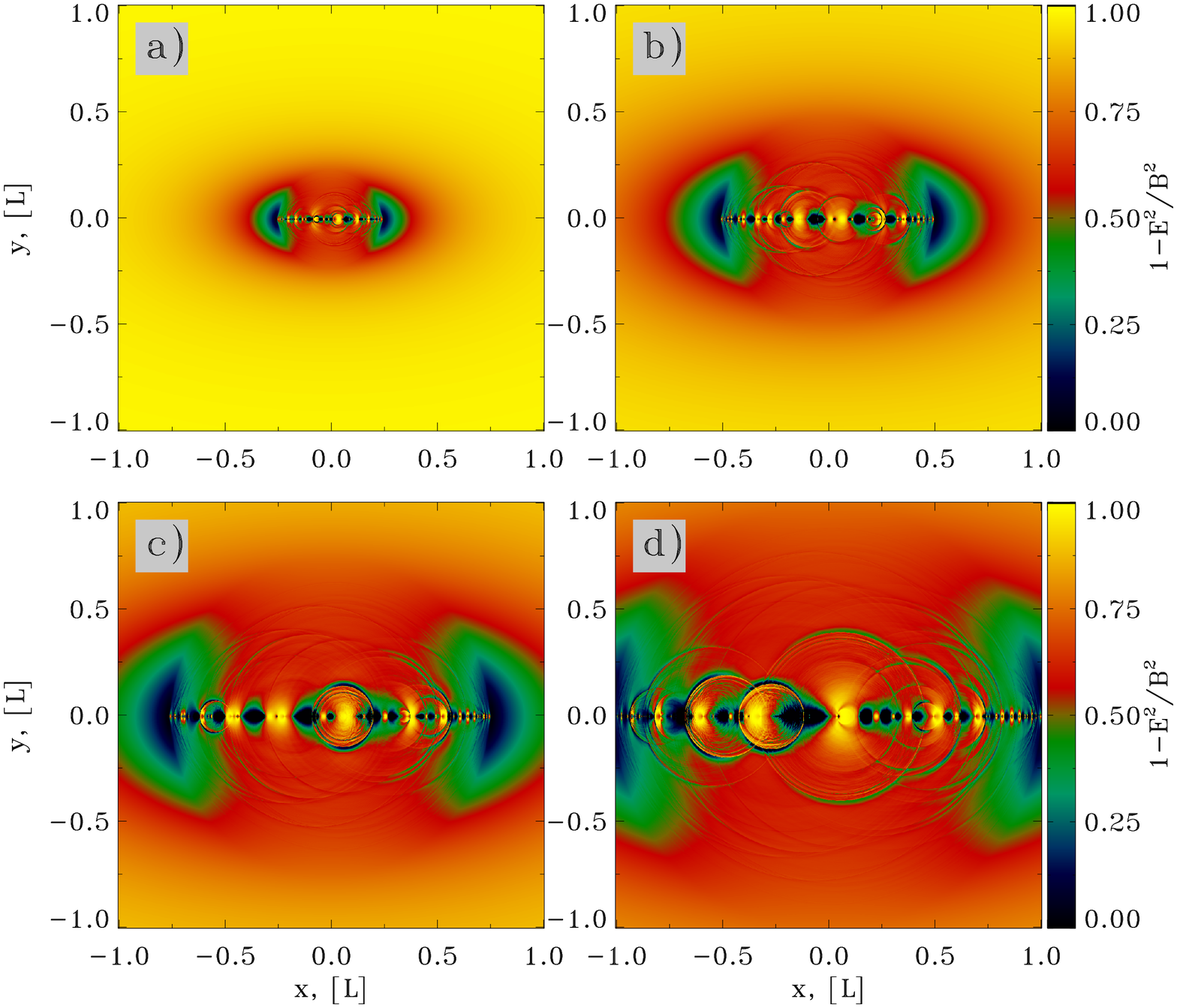} 
\includegraphics[width=.49\textwidth]{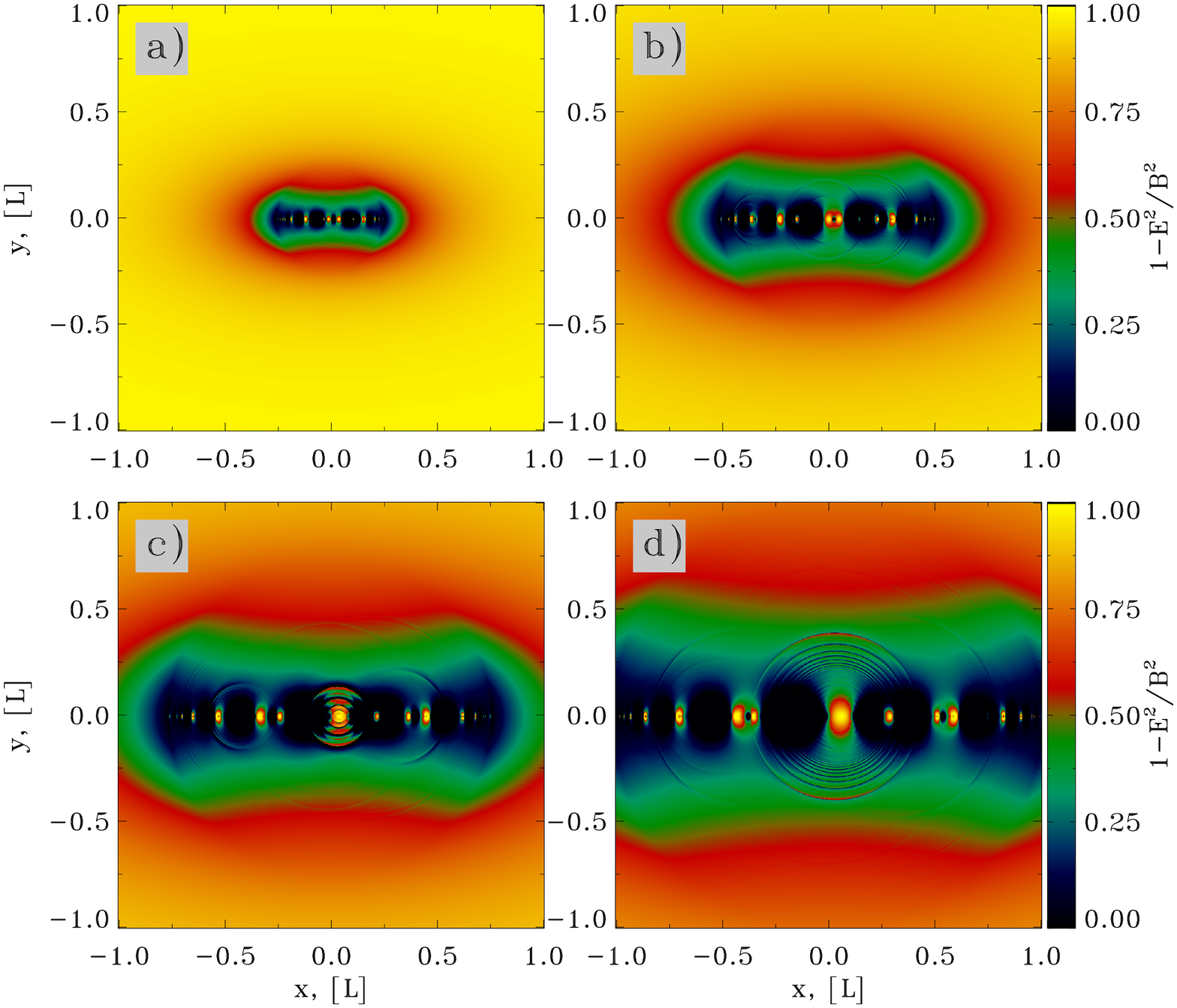} 
\caption{Initial phase of a solitary X-point collapse in PIC simulations with zero guide field, for two different magnetizations: $\sigma_L =4\ex{3}$  (left)  and  $\sigma_L =4\ex{4}$  (right). The plots show the quantity $1-E^2/B^2$ (strictly speaking, we plot $\max[0,1-E^2/B^2]$) at $ct/L=$0.25, 0.5, 0.75 and 1, from panel (a) to (d).}
\label{sigma3} 
\end{figure}

 \begin{figure}[!ht]
 \centering
\includegraphics[width=.49\textwidth]{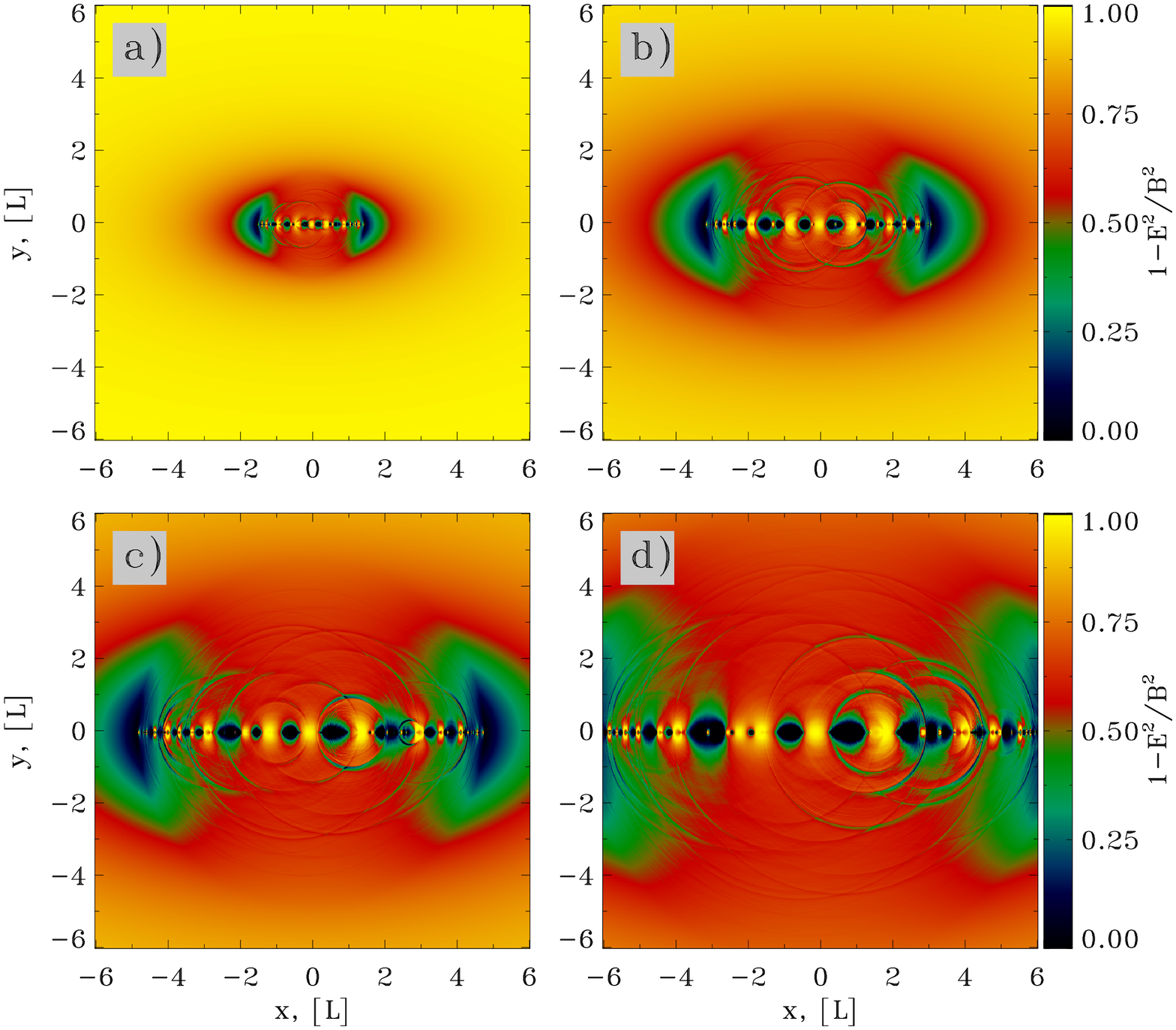} 
\includegraphics[width=.49\textwidth]{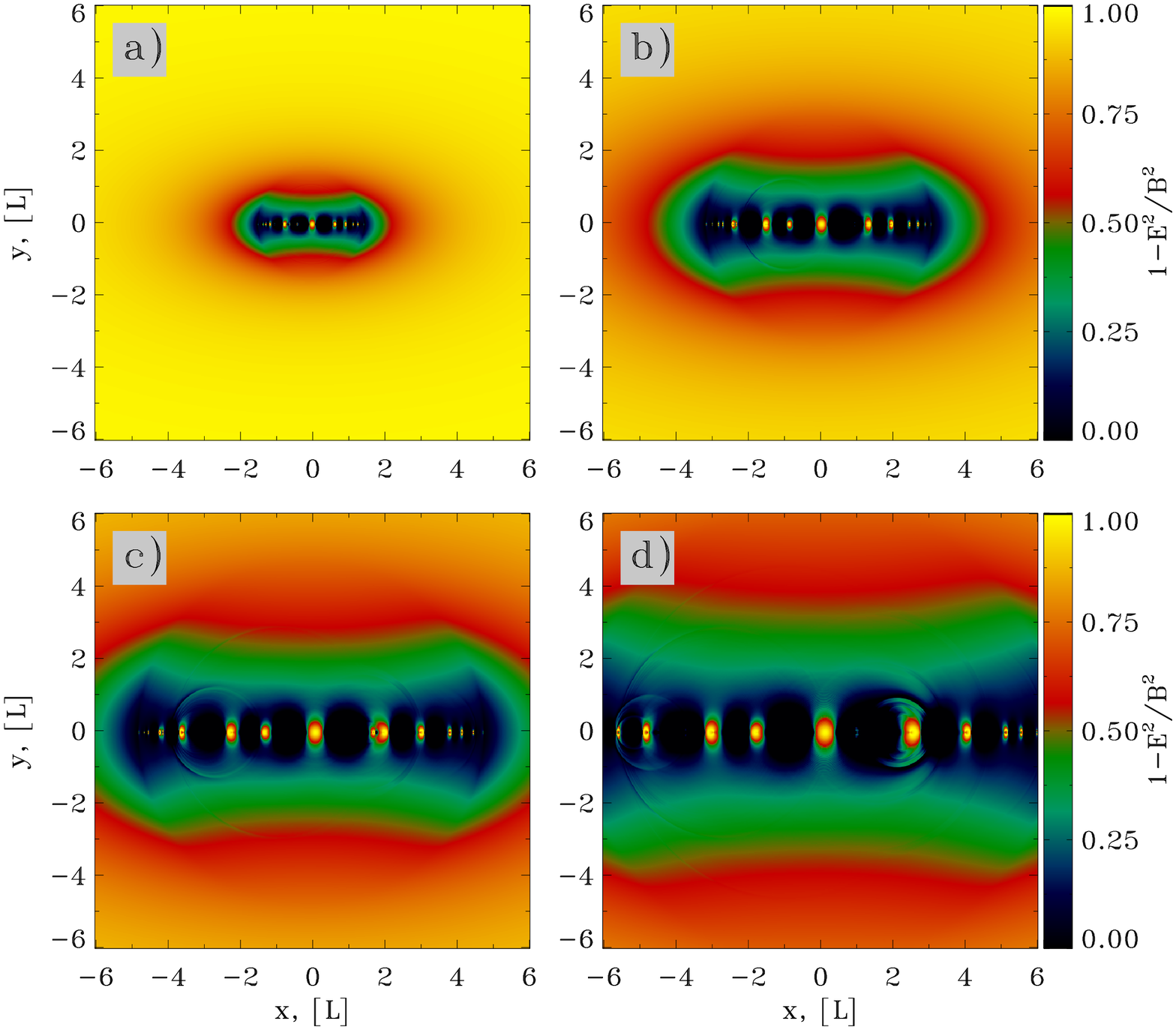} 
\caption{Late time evolution of the X-point collapse in PIC simulations with zero guide field, for two different magnetizations: $\sigma_L =4\ex{3}$  (left)  and  $\sigma_L =4\ex{4}$  (right). The plots show the quantity $1-E^2/B^2$ (strictly speaking, we plot $\max[0,1-E^2/B^2]$) at $ct/L=$1.5, 3, 4.5, 6, from panel (a) to (d).}
\label{sigma3-1} 
\end{figure}
%%%%%%%%%%%%%%%%%%%
 \begin{figure}[!ht]
 \centering
\includegraphics[width=.49\textwidth]{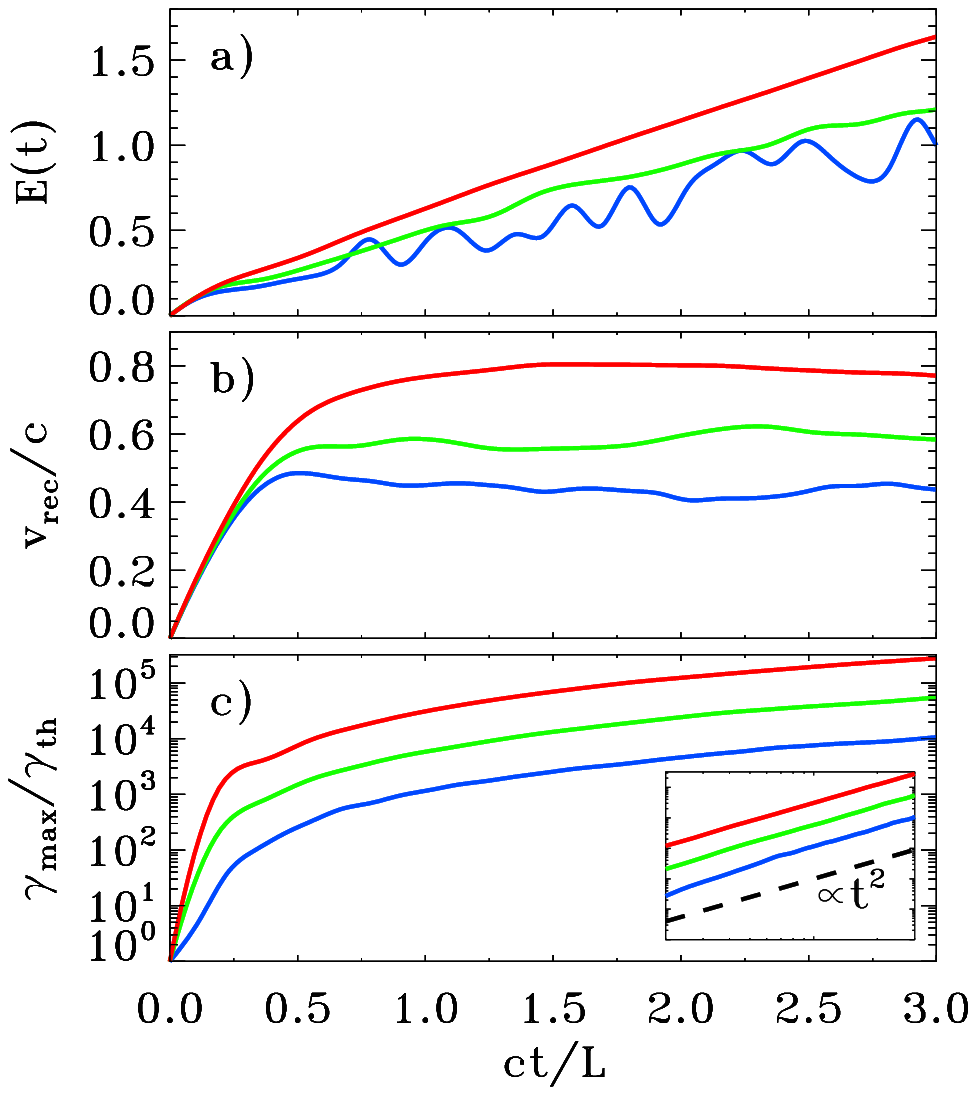} 
\caption{Temporal evolution of various quantities from PIC simulations of solitary X-point collapse with zero guide field, for three values of the magnetization: $\sigma_L =4\times10^2$  (blue), $\sigma_L =4\ex{3}$  (green)  and  $\sigma_L =4\ex{4}$ (red). The corresponding plot for the case of nonzero guide field is in \fig{xguidetimecomp}. As a function of time, we plot: (a) the value of the electric field strength $E(t)$ at the location $(-0.1L,0.1L)$ in units of the initial magnetic field at $x=L$; (b), the reconnection rate, defined as the  inflow speed along the $y$ direction averaged over a square of side equal to $L$ around the central region; (c) the maximum particle Lorentz factor $\gamma_{\rm max}$ (as defined in \eq{ggmax}), in units of the thermal value $\gamma_{\rm th}\simeq 1+(\hat{\gamma}-1)^{-1} kT/m c^2$, which in this case of a cold plasma reduces to $\gamma_{\rm th}\simeq 1$; the inset in panel (c) shows the same quantity on a double logarithmic scale, demonstrating that $\gamma_{\rm max}\propto t^2$ (black dashed line).}
\label{fig:xtimecomp} 
\end{figure}

\subsection{Stressed X-point collapse without guide field}\label{sec:xnoguide}

Figure~\ref{sigma3} shows the initial phase  ($ct/L\leq 1$) of the collapse of an X-point in PIC simulations with $\lambda=1/\sqrt{2}$ and with zero guide field,  for two different 
magnetizations: $\sigma_L =4\ex{3}$  (left)  and  $\sigma_L =4\ex{4}$  (right). We plot the quantity $1-E^2/B^2$ (more precisely, we present $\max[0,1-E^2/B^2]$), in order to identify the regions where the electric field is comparable to the magnetic field (green or blue areas in the plot). For both  $\sigma_L =4\ex{3}$  (left)  and  $\sigma_L =4\ex{4}$  (right), the current sheet is subject to copious fragmentation into plasmoids since early times, in contrast with the guide-field case (compare with Fig.~\ref{x-inner-pic}). There, the current sheet was supported by the pressure of the compressed guide field, and therefore its thickness was larger, making it less prone to fragmentation (at a fixed time $ct/L$). In addition, a comparison of Fig.~\ref{sigma3}, which refers to early times (up to $ct/L=1$), with  Fig.~\ref{sigma3-1}, that follows the system up to $ct/L=6$, shows the remarkable self-similarity of the evolution, for both magnetizations.  {First, the macroscopic distribution of $E^2/B^2$ (and hence that of the drift velocity) at later times is a scaled copy of that at previous times, with the overall length scale increasing linearly with time (at the speed of light). This implies that the reconnection rate over the whole configurations remains fixed in time, as we indeed confirm below.  Second, the size of the largest plasmoids generated in the current sheet is also a fixed fraction of the overall scale, $\sim 0.1$ of the current sheet length.} 

In  Figs.~\ref{sigma3} and \ref{sigma3-1}, the plasmoids generated  by the secondary tearing instability \citep{uzdensky_10} appear as yellow structures, i.e., with magnetic energy much larger than the electric energy. In contrast, the region in between each pair of plasmoids harbors a {\it microscopic} X-point, where the electric field can exceed the magnetic field. The size of these  {microscopic X-points is controlled by plasma kinetics, in contrast to the original macroscopic X-point. } 
They play a major role for particle injection into the acceleration process, as we argue in the next subsection.

As observed in the case of guide-field collapse, the two bow-shaped regions ahead of the Y-points (to the left and to the right of the reconnection layer) are moving relativistically, yielding $E\sim B$ (green and blue colors in the figures). In addition, in the high-magnetization case  $\sigma_L=4\times10^4$ (right side of Figs.~\ref{sigma3} and \ref{sigma3-1}), a {\it macroscopic} region appears in the bulk inflow where the electric field is comparable to the magnetic field. Here, the inflow rate approaches the speed of light, as we have already described in the case of guide-field reconnection (right side of Fig.~\ref{pic-xcg-bme}).

This is further illustrated in \fig{xtimecomp}, where we present the temporal evolution of various quantities, from a suite of PIC simulations of X-point collapse with vanishing guide field, having three different magnetizations: $\sigma_L =4\times10^2$  (blue), $\sigma_L =4\ex{3}$  (green)  and  $\sigma_L =4\ex{4}$ (red). The reconnection rate $v_{\rm rec}/c$ (panel (b)), which is measured as the inflow speed averaged over a {\it macroscopic} square of side equal to $L$ centered at $x=y=0$, shows in the asymptotic state (i.e., at $ct/L\gtrsim 0.5$) a clear dependence on $\sigma_L$, reaching $v_{\rm rec}/c\sim 0.8$ for our high-magnetization case $\sigma_L =4\ex{4}$ (red). This trend has already been described by \citet{liu_15}. The critical difference, though, is that their measurement was performed on {\it microscopic} skin-depth scales, whereas our results show that  reconnection velocities near the speed of light can be achieved over {\it macroscopic} scales $L\gg\comp$. In addition, such inflow speed is significantly larger than what is measured on macroscopic scales in the case of plane-parallel steady-state reconnection, where the reconnection rate typically approaches $v_{\rm rec}/c\sim 0.2$ in the high-magnetization limit \citep[][]{2014ApJ...783L..21S,2015ApJ...806..167G}.  

The dependence of the reconnection speed on $\sigma_L$ is also revealed in \fig{xtimecomp}(a), where we present the temporal evolution of the electric field $E(t)$ measured at $(x,y)=(-0.1L,0.1L)$, in units of the initial magnetic field at $x=L$. The variation in slope in \fig{xtimecomp}(a) is indeed driven by the different reconnection speeds, since $E\sim v_{\rm rec} B/c$ in the inflow region. 

{Interestingly, the electric field increases linearly with time. This is ultimately a manifestation of the self-similar macroscopic evolution of the system.  Indeed,  since in the initial configuration the magnetic field strength grows linearly with distance from the origin (i.e., the center of the X-point) and the current sheet size grows linearly with time, the mean magnetic and electric fields in the volume surrounding the current sheet must also grow linearly, with their scaled distributions unchanged. }
The temporal evolution of the electric field has a direct impact on the maximum particle energy, which is shown  in \fig{xtimecomp}(c). Quite generally, its time evolution will be
\be\label{eq:ggmax2}
\gammamax \propto E t\propto v_{\rm rec} B t
\ee
Since both $E$ and $B$ in the reconnection region are scaling linearly with time  (see \fig{xtimecomp}(a)), one expects $\gamma_{\rm max}\propto t^2$, as indeed confirmed by the inset of panel (c) (compare with the dashed black line). This scaling is faster than in plane-parallel steady-state reconnection, where $E(t)$ is constant in time, leading to $\gamma_{\rm max}\propto t$. From the scaling in \eq{ggmax2}, one can understand the different normalizations of the curves in \fig{xtimecomp}(c). {Since $B\propto \sqrt{\sigma_L}$ we find that at fixed time 
$$ 
\gammamax\propto v_{\rm rec} \sqrt{\sigma_L}
$$ 
and hence it grows with the magnetization. We find that for the model with $\sigma_L=4\times 10^4$ the reconnection rate is
about twice that of the model with $\sigma_L=4\times 10^2$ (panel (b) in \fig{xtimecomp}) and hence $\gammamax$ should be 20 times higher. }
This is in excellent agreement with the data in \fig{xtimecomp}(c) (compare blue and red curves).

%%%%%%%%%%%%%%%%%%%
 \begin{figure}[!ht]
 \centering
\includegraphics[width=.49\textwidth]{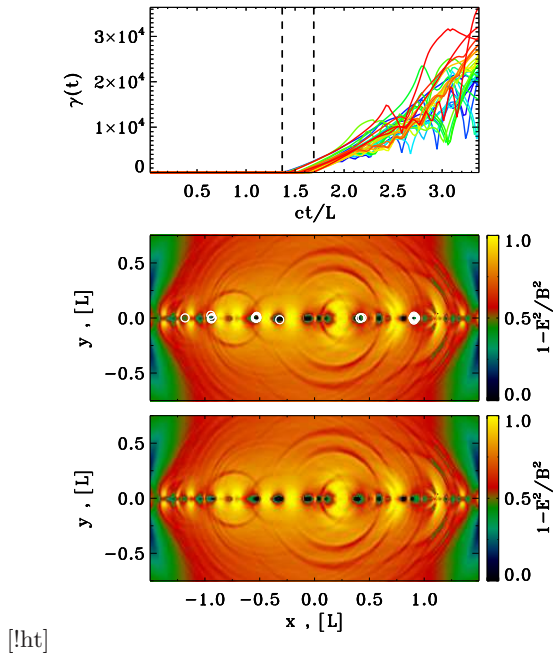} 
\caption{Physics of particle injection into the acceleration process, from a PIC simulation of stressed X-point collapse with vanishing guide field and $\sigma_L=4\times10^2$. Top panel: we select all the particles that exceed the threshold $\gamma_{\rm 0}=30$ within a given time interval (chosen to be $1.4\leq ct_0/L\leq1.7$, as indicated by the vertical dashed lines), and we plot the temporal evolution of the Lorentz factor of the 30 particles that at the final time reach the highest energies. The particle Lorentz factor increases as $\gamma\propto t^2-t_0^2$, where $t_0$ marks the onset of acceleration (i.e., the time when $\gamma$ first exceeds $\gamma_{\rm 0}$). Middle panel: for the same particles as in the top panel, we plot their locations at the onset of acceleration with open white circles, superimposed over the 2D plot of $1-E^2/B^2$ (more precisely, of $\max[0,1-E^2/B^2]$) at $ct/L=1.55$. Comparison of the middle panel with the bottom panel shows that particle injection is localized in the vicinity of the X-points in the current sheet (i.e., the blue regions where $E>B$).}
\label{fig:xaccfluid} 
\end{figure}
 \begin{figure}[!ht]
 \centering
\includegraphics[width=.49\textwidth]{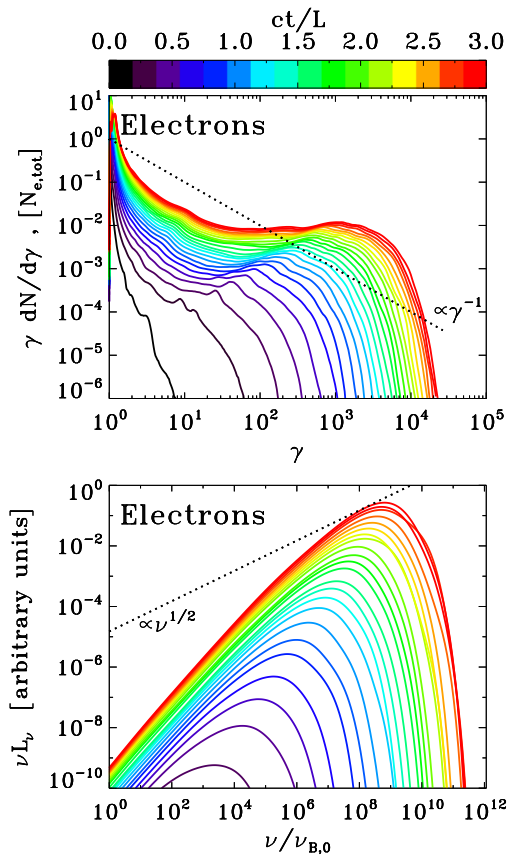} 
\caption{Particle energy spectrum and synchrotron spectrum from a PIC simulation of stressed X-point collapse with vanishing guide field and $\sigma_L=4\times10^2$. Time is measured in units of $L/c$, see the colorbar at the top. Top panel: evolution of the electron energy spectrum normalized to the total number of electrons. At late times, the spectrum approaches a hard distribution $\gamma dN/d\gamma\propto {\rm const}$, much harder than the dotted line, which shows the case $\gamma dN/d\gamma\propto \gamma^{-1}$ corresponding to equal energy content in each decade of $\gamma$. Bottom panel: evolution of the angle-averaged synchrotron spectrum emitted by electrons. The frequency on the horizontal axis is in units of $\nu_{B,0}=\sqrt{\sigma_L}\omega_{\rm p}/2\pi$. At late times, the synchrotron spectrum approaches a power law with $\nu L_\nu\propto \nu$, which just follows from the fact that the electron spectrum is $\gamma dN/d\gamma\propto {\rm const}$. This is much harder than the dotted line, which indicates the slope $\nu L_\nu\propto \nu^{1/2}$ resulting from an electron spectrum $\gamma dN/d\gamma\propto \gamma^{-1}$ (dotted line in the top panel).}
\label{fig:xspec} 
\end{figure}
 \begin{figure}[!ht]
 \centering
\includegraphics[width=.49\textwidth]{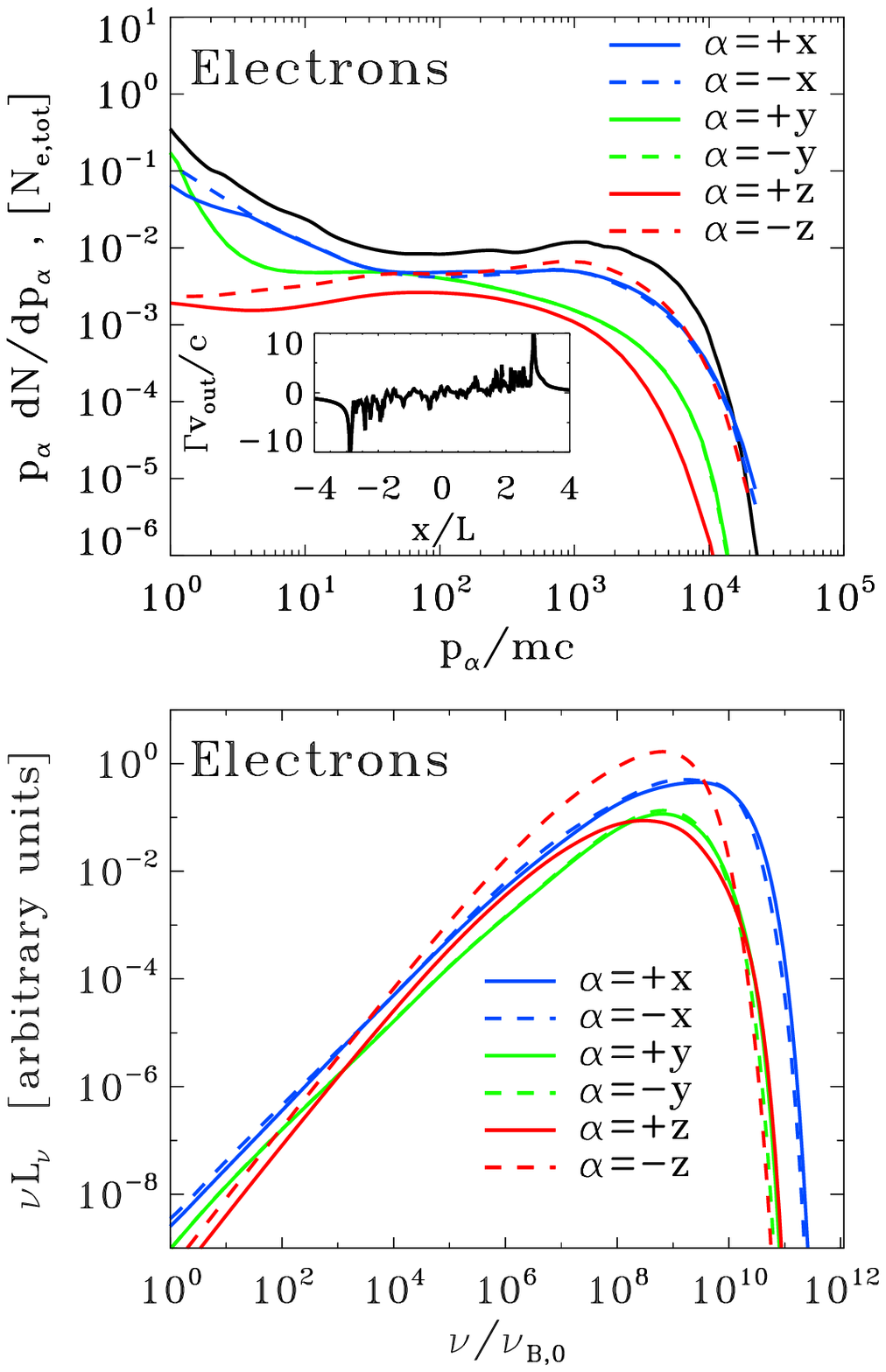} 
\caption{Particle momentum spectrum and anisotropy of the synchrotron spectrum from a PIC simulation of stressed X-point collapse with vanishing guide field and $\sigma_L=4\times10^2$. Top panel: electron momentum spectrum at the final time $ct/L=3$ along different directions, as indicated in the legend. The total momentum spectrum (i.e., independent of direction) is indicated with a solid black line for comparison. The highest energy electrons are beamed along the direction $x$ of the reconnection outflow (blue lines) and along the direction $-z$ of the accelerating electric field (red dashed line; positrons will be beamed along $+z$, due to the opposite charge).  The inset shows the 1D profile along $x$ of the bulk four-velocity in the outflow direction (i.e., along $x$), measured at $y=0$.  Bottom panel: synchrotron spectrum at the final time $ct/L=3$ along different directions (within a solid angle of $\Delta \Omega/4\pi\sim 3\times10^{-3}$), as indicated in the legend. The resulting  anisotropy of the synchrotron emission is consistent with the particle anisotropy illustrated in the top panel.}
\label{fig:xspecmom} 
\end{figure}
 \begin{figure}[!ht]
 \centering
\includegraphics[width=.49\textwidth]{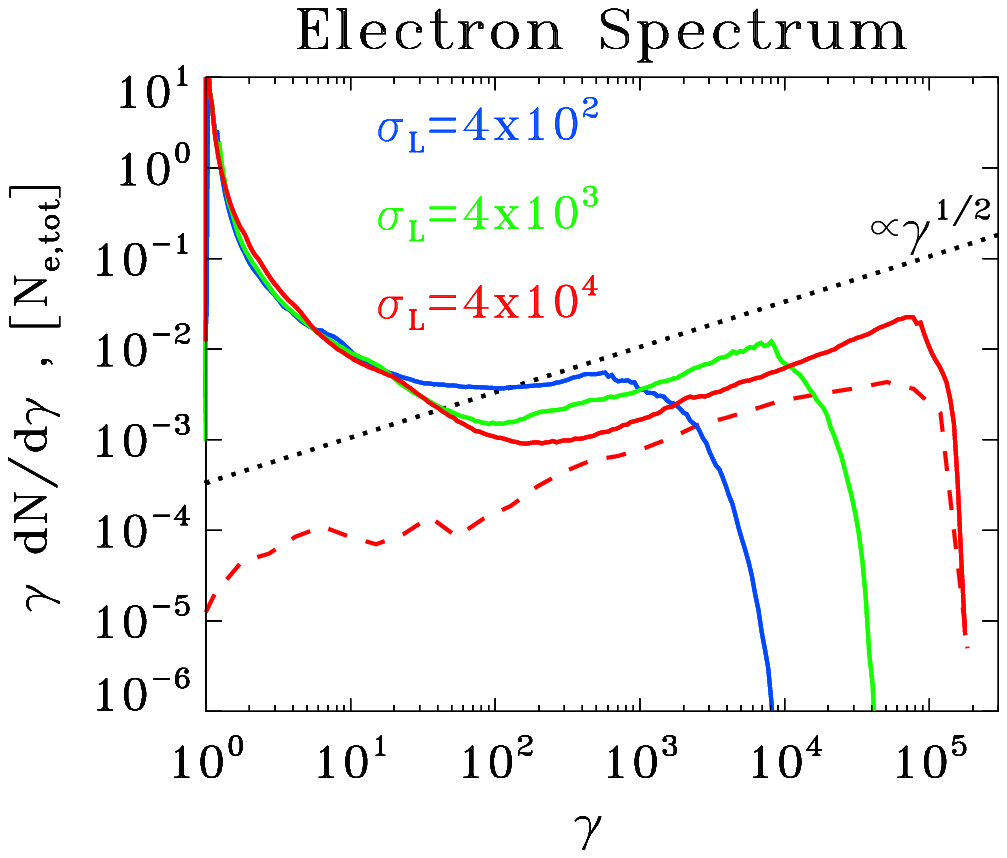} 
\caption{{Dependence of the electron energy spectrum on the magnetization,  for three values of $\sigma_L$ (same values and color coding as in \fig{xtimecomp}) and vanishing guide field: $\sigma_L =4\times10^2$  (blue), $\sigma_L =4\ex{3}$  (green)  and  $\sigma_L =4\ex{4}$ (red). All the spectra are computed at $ct/L=1.8$. At high magnetizations, two components can be seen in the spectrum: a steep low-energy component and a hard high-energy population that can be fitted as $\gamma dN/d\gamma\propto \gamma^{1/2}$ (black dotted line). The red dashed line is the particle spectrum for $\sigma_L =4\ex{4}$ at the same time as the red solid line, but including only the particles located in regions where $E>B$.}}
\label{fig:xspeccomp} 
\end{figure}

\subsection{Particle acceleration and emission signatures}\label{sec:xaccel}
In this section, we explore the physics of particle acceleration in a stressed X-point collapse with vanishing guide field, and we present the resulting particle distribution and synchrotron emission spectrum. In \fig{xaccfluid}, we  follow the trajectories of a number of particles in a simulation with $\sigma_L=4\times10^2$. The particles are selected such that their Lorentz factor exceeds a given threshold $\gamma_0=30$ within the time interval $1.4\leq ct_0/L\leq1.7$, as indicated by the vertical dashed lines in the top panel. The temporal evolution of the Lorentz factor of such particles, presented in the top panel for the 30 positrons reaching the highest energies, follows the track $\gamma\propto t^2-t_0^2$ that is expected from $d\gamma/dt\propto E(t)\propto t$. Here, $t_0$ is the injection time, when the particle Lorentz factor $\gamma$ first exceeds the threshold $\gamma_{\rm 0}$. The individual histories of single positrons might differ substantially, but overall the top panel of \fig{xaccfluid} suggests that the acceleration process is dominated by direct acceleration by the reconnection electric field, as indeed it is expected for our configuration of a large-scale stressed X-point \citep[see][for a discussion of acceleration mechanisms in plane-parallel reconnection]{2014ApJ...783L..21S,2015ApJ...806..167G,nalewajko_15}. We find that the particles presented in the top panel of \fig{xaccfluid} are too energetic to be confined within the small-scale plasmoids in the current sheet, so any acceleration mechanism that relies on plasmoid mergers is found to be unimportant, in our setup.

Particle injection into the acceleration process happens in the charge-starved regions where $E>B$, i.e., in the small-scale X-points that separate each pair of secondary plasmoids in the current sheet. Indeed, for the same particles as in the top panel, the middle panel in \fig{xaccfluid} presents their locations at the onset of acceleration with open white circles, superimposed over the 2D plot of $1-E^2/B^2$ (more precisely, of $\max[0,1-E^2/B^2]$). Comparison of the middle panel with the bottom panel shows that particle injection is localized in the vicinity of the small-scale X-points in the current sheet (i.e., the blue regions where $E>B$). Despite occupying a relatively small fraction of the overall volume, such regions are of paramount importance for particle acceleration.

The temporal evolution of the electron energy spectrum is presented in the top panel of \fig{xspec}, for a simulation with $\sigma_L=4\times 10^2$. As the spectral cutoff grows as $\gammamax\propto t^2$ (see also the inset in \fig{xtimecomp}(c)), the spectrum approaches a hard power law $\gamma dN/d\gamma\propto \rm const$. The measured spectral slope is consistent with the asymptotic power-law index obtained in the limit of high magnetizations from PIC simulations of relativistic plane-parallel reconnection \citep[][]{2014ApJ...783L..21S,2015ApJ...806..167G,2016ApJ...816L...8W}. Due to energy conservation, such hard slopes would not allow the spectrum to extend much beyond the instantaneous value of the magnetization just upstream of the current sheet (as we have explained before, in our setup the magnetization at the current sheet increases quadratically with time, since $B(t)\propto t$), in line with the arguments of \citet{2014ApJ...783L..21S} and \citet{2016ApJ...816L...8W}. 

{However, we find evidence for a possible solution of this ``energy crisis.'' In \fig{xspeccomp}, we explore the dependence of the particle energy spectrum on the magnetization at $ct/L=1.8$, in the case of vanishing guide field. We find that at high $\sigma_L$ the population of accelerated particles is composed of two components, separated by a break energy (for the red solid curve corresponding to $\sigma_L=4\times10^4$ in \fig{xspeccomp}, the break occurs at a Lorentz factor $\gamma\sim 200$): a low-energy soft component, whose spectral slope is slightly steeper than $\gamma dN/d\gamma\propto \gamma^{-1}$ (corresponding to equal energy content per decade);\footnote{The power-law nature of the low-energy component is clearly apparent in the momentum spectrum, see the black line in the top panel of  \fig{xspecmom}.} and a high-energy hard population, with $\gamma dN/d\gamma\propto \gamma^{1/2}$ (so, with the highest energy particles dominating both the number and the energy census). The presence of the soft component, whose energy per particle is significantly lower than the average energy (namely, of the local magnetization), allows the high-energy component to stretch to higher energies, potentially offsetting the energy crisis.}

{The two sub-populations have different origins: by tracking individual particles, we find that the soft component is dominated by particles belonging to secondary plasmoids, that are accelerated during plasmoid mergers; in contrast, the hard component is populated by particles that are accelerated by the strong electric fields of the charged-starved regions with $E>B$, and are nearly unaffected by the presence of secondary plasmoids. In fact, the spectrum of the particles located in $E>B$ regions, presented in \fig{xspeccomp} with a dashed red line, only shows the hard component.}

{Our simulations of  X-point collapse in the presence of a nonzero guide field provide further support to this conclusion. At early times, when no secondary plasmoids are present (in fact, the guide field suppresses the secondary tearing mode), particle energization can only occur via direct acceleration by the charge-starved electric fields in $\bmath{E}\cdot\bmath{B}\neq0$ regions. At these times, only the hard component with $\gamma dN/d\gamma\propto \gamma^{1/2}$ appears in the corresponding particle spectrum (not shown). At later times, when the guide field at the current sheet becomes sub-dominant with respect to the in-plane fields, secondary plasmoids can be generated, and an additional soft component appears in the particle spectrum.}

{We conclude by analyzing the anisotropy of the particle distribution, for $\sigma_L=4\times10^2$.} In the top panel of \fig{xspecmom}, we plot the electron momentum spectrum at the final time $ct/L=3$ along different directions, as indicated in the legend. The particle distribution is significantly anisotropic. The highest energy electrons are beamed along the direction $x$ of the reconnection outflow (blue lines) and along the direction $-z$ of the accelerating electric field (red dashed line; positrons will be beamed along $+z$, due to the opposite charge). This is consistent with earlier PIC simulations of plane-parallel reconnection in a small computational box, where the X-point acceleration phase was still appreciably imprinting the resulting particle anisotropy \citep{2012ApJ...754L..33C,2013ApJ...770..147C,cerutti_13b,kagan_16}. In contrast, plane-parallel reconnection in larger computational domains generally leads to quasi-isotropic particle distributions \citep{2014ApJ...783L..21S}. In our setup of a large-scale X-point, we would expect the same level of strong anisotropy measured in small-scale X-points of plane-parallel reconnection, as indeed demonstrated in the top panel of \fig{xspecmom}. Most of the anisotropy is to be attributed to the ``kinetic beaming''  discussed by \citet{2012ApJ...754L..33C}, rather than beaming associated with the bulk motion (which is only marginally relativistic, see the inset in the top panel of \fig{xspecmom}).

The angle-averaged synchrotron spectrum expected from a relativistic X-point collapse is shown in the bottom panel of \fig{xspec}. For each macro-particle in our PIC simulation, we compute the instantaneous radius of curvature and the corresponding synchrotron emission spectrum. We neglect synchrotron cooling in the particle equations of motion \citep[unlike][]{2013ApJ...770..147C,cerutti_13b,kagan_16}, and we do not consider the effect of synchrotron self-absorption and the Razin suppression at low frequencies. At late times, the synchrotron spectrum approaches a power law with $\nu L_\nu\propto \nu$, which just follows from the fact that the electron spectrum is $\gamma dN/d\gamma\propto {\rm const}$. This is consistent with the spectrum of the Crab flares. The frequency on the horizontal axis is in units of $\nu_{B,0}=\sqrt{\sigma_L}\omega_{\rm p}/2\pi$. Given the maximum particle energy in the top panel, $\gammamax\sim 10^4$, one would expect the synchrotron spectrum to cut off at $\nu_{\rm max}\sim \gamma_{\rm max}^2\nu_{B,0}\sim 10^8 \nu_{B,0}$, as indeed confirmed in the bottom panel. The bottom panel of \fig{xspecmom} shows the synchrotron spectrum at the final time $ct/L=3$ along different directions (within a solid angle of $\Delta \Omega/4\pi\sim 3\times10^{-3}$), as indicated in the legend. The resulting  anisotropy of the synchrotron emission is consistent with the particle anisotropy illustrated in the top panel of \fig{xspecmom}. In addition, one can see that the resulting synchrotron spectrum along the direction $-z$ of the accelerating electric field (dashed red line) appears even harder than the spectrum along $x$ or $y$.

%% file: conclusion1.tex
\section{Discussion and Conclusions}
\label{Conc1}

In highly magnetized plasma, the collapse of a uniformly compressed X-point proceeds very rapidly, on the light-crossing time of the configuration. As a result, the collapse cannot be described using classical quasi-static approach and the generated electric field is of the order of the magnetic one. In the framework of force-free electrodynamics, we find that without guide field, the x-point immediately develops a singularity -- collapses into a current sheet  bounded by two Y-points, which fly away at the speed of light. With guide field, the development of singularities is delayed and the initial phase of the collapse can be described analytically. We have shown that for the central region of the X-point, where the guide field strength exceeds that of the in-plane magnetic field, the problem reduces to a single ODE for the squeeze parameter and we have found a simple asymptotic solution to this equation. The solution describes a systematic increase of this parameter, thus indicating that in highly magnetized plasma X-points remain unstable to collapse.  This conclusion is confirmed by the results of computer simulations, both force-free and PIC.     

The force-free and PIC simulation agree very well until the point where the development of strong electric field leads to a breakdown of the force-free approximation near the plane of the collapse. Before this point, the plasma kinetic effects are weak and the evolution is totally controlled by the large-scale magnetic stresses. After this point, the kinetic effects become increasingly important in determining the current sheet structure and its feedback to the surrounding electromagnetic field.  Although qualitatively, and in many ways quantitatively, the force-free and PIC solutions remain similar, a number of differences become manifest. E.g. in the PIC simulations the current sheet develops magnetic islands (plasmoids) whereas the force-free current sheet remains rather featureless\footnote{Although such islands may form in force-free simulations, at least for some types of current sheets, this process is governed by numerical factors and hence not physical. }.  

As the current sheet expands into the region where the guide field is week, its evolution becomes self-similar. 
This is observed both in the force-free and PIC simulations. Without guide field, the transition to this regime is very quick. 
In PIC simulations this occurs as soon as the length of the current sheet significantly exceeds the skin depth. In the self-similar regime, the macroscopic distribution of the magnetic and electric fields around the current sheet remains unchanged but the field strengths scale with the size of the current sheet, which increases at almost the speed of light. Hence the strength of  average electric and magnetic fields in the reconnecting region grows linearly with time, whereas the overall reconnection rate remains unchanged.          

{We find that as $\sigma$ increases the reconnection rate approaches the speed of light on a macroscopic scale. This is in contrast with the previous PIC simulations of magnetic reconnection where such high rates have been seen only on the microscopic scales. The large macroscopic stresses typical for the collapsing X-point configuration appear to be the main factor behind the increase.  As the strength of the magnetic field brought into the reconnection zone grows linearly with time, so does the strength of the reconnection electric field. This allows the maximum energy of accelerated particles to increase as $\propto t^2$.} 

{Particle acceleration is a self-consistent by-product of the X-point collapse. Regardless of whether or not a guide field is present in the initial configuration, the highest energy particles are injected into the acceleration process in charged-starved regions (i.e., where $\bmath{E}\cdot\bmath{B}\neq0$ in the case with guide field, or where $E>B$ for  zero guide field), and energized via direct acceleration by the reconnection electric field.}  {As a result, the maximum particle energy does increase as $\propto t^2$.}  {While secondary plasmoids are continuously generated in the current sheet (in the cases with guide field, at sufficiently late times), in our setup they are not instrumental for the acceleration of the highest energy particles. The high-energy part of the spectrum is hard, with power-law slope even harder than $-1$, and populated by highly anisotropic particles, beamed primarily along the direction of the accelerating electric field.}

{The hard high-energy tail can be problematic if it includes a large fraction of the accelerated particles as in this case the particle energy is limited by the mean magnetization $\sigma$ of plasma brought into the reconnection zone. This would require unrealistic 
$\sigma\approx 10^{10}$ in order to explain Crab's flares.  However, as soon as  the secondary plasmoids are formed in the current sheet, we observe an emergence of the second population of the accelerated particles. These particles are trapped inside the plasmoids, they are accelerated mostly during plasmoid mergers and their spectrum is significantly softer. By the end of our simulations, the number of particles trapped in the plasmoids exceeds that of the hard-energy tail by several orders of magnitude, thus indicating the possible route of resolving this kind of $\sigma$-problem. Unfortunately, due to the computational limitations we have not been able to reach the particle energies $\gamma\approx 10^{10}$ 
typical for the Crab flares. Additional studies are needed to clarify this issue. }

% {\bf (Lorenzo, does this make sense?  If it does, do we have a trend  for the number ratio of two populations against the maximum Lorentz factor?  )}}

Since magnetic X-points are unstable to collapse, this brings about the question of how they can be formed in the first place. 
We cannot exclude that static macroscopic configurations of the sort can be maintained  in controlled laboratory experiments, but they are most unlikely to be found in highly dynamic conditions of astrophysical plasma. Here we studied this configuration as an example of a system where large scale magnetic stresses drive magnetic reconnection and dictate its rate. In the second paper of the series, we consider another artificial example of initially steady-state configuration, the so-called ABC magnetic structure.  This periodic configuration has local macroscopic  X-points. We find that this configuration is unstable and that the development of this instability triggers collapse of these X-points in the similar fashion to what we described here.  This result suggests that highly-magnetized plasma configurations are generally unstable and exist mainly in the state of rapid restructuring on the light-crossing time. During this restructuring, macroscopic stresses drive magnetic reconnection, causing local magnetic dissipation  and acceleration of non-thermal particles. In the final paper of the series, we no longer consider a static initial configuration but study a collision of magnetic current tubes. Such a collision is also accompanied by  development of X-points (highly compressed ones)  and also leads to magnetic reconnection driven by macroscopic magnetic stresses. Thus, it seems that we are dealing with a rather general phenomenon which may have important applications in relativistic astrophysics.

%% file: appendix1.tex
 
%ssssssssssssssssssssssssssssssssssssssssssssss
\section{Stability of  unstressed X-point}
\label{un-stressed}
%ssssssssssssssssssssssssssssssssssssssssssssss

The above analytical solution shows that the steady-state X-point solution is kind of 
unstable. This rises the question of how such configuration can be created in a first place. 
Indeed, unstable states of a dynamical system cannot be reached via its natural evolution.  
However, the X-point configuration considered in this analyses occupies the whole space and 
so is the perturbation that leads to its collapse. In reality, X-points and their perturbations 
occupy only finite volume and in order to address the stability issue comprehensively 
one has to study finite size configurations. 

\begin{figure}[h!]
\centering
\includegraphics[width=.35\textwidth]{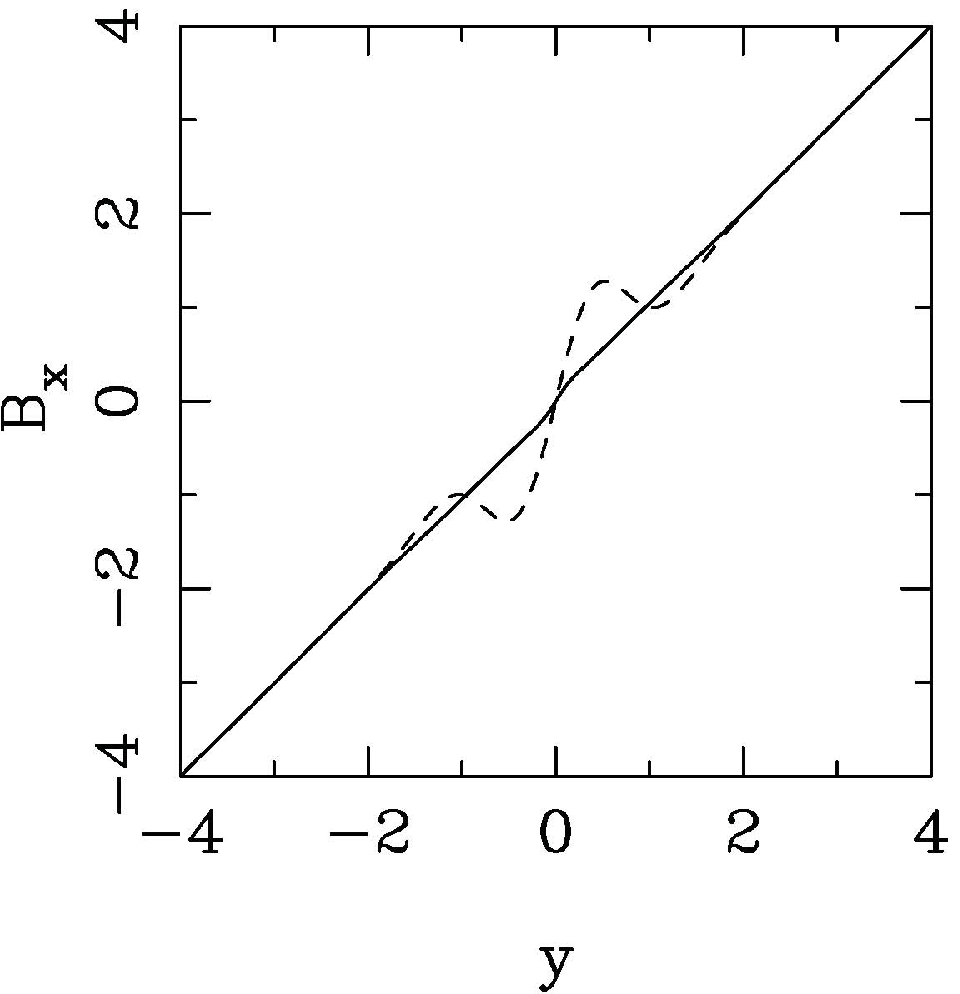}
\includegraphics[width=.35\textwidth]{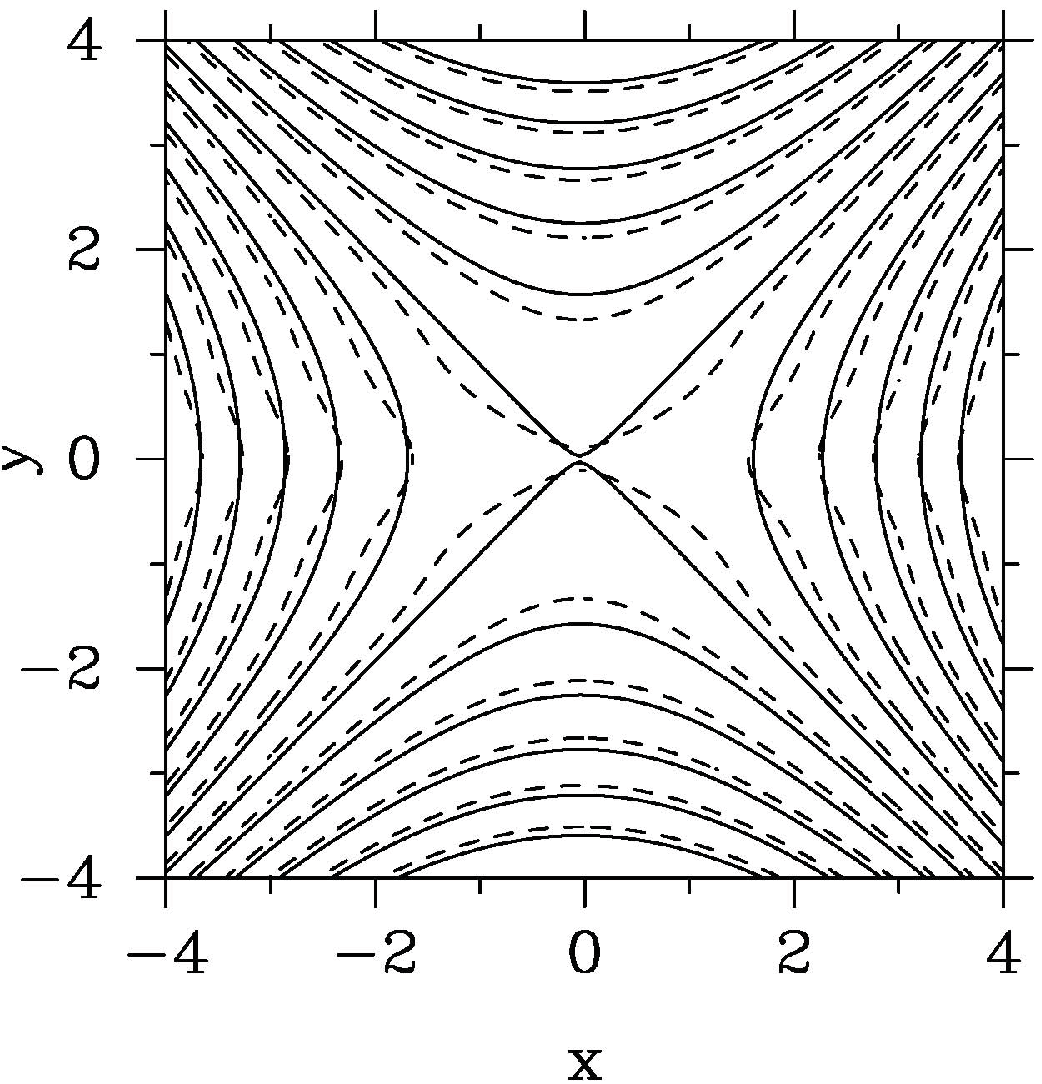}
\caption{Stability of the X-point to small-scale perturbations in force-free simulations.
The left panel shows the x component of the magnetic field 
along the line $y=0$. The dashed line corresponds to the initial perturbed solution.
The solid line corresponds to the numerical solution at $t=7$. 
The right panel shows the magnetic field lines of the initial solution (dashed lines) and 
the solution at $t=7$. 
}
\label{stab}
\end{figure}

In this section we describe the response of X-point to small-scale perturbations studied 
via force-free simulations. In one of our experiments, we perturbed the steady-state X-point
configuration by varying only the x-component of the magnetic field:

$$
   \delta B_x= B_\perp \sin(\pi y/L) \exp(-(y/L)^2)\,.
$$ 
Obviously, the length scale this perturbation is $L$ and to ensure that it is small we 
select a computational domain whose size is much larger than $L$. In this particular case we 
put $B_\perp=L=1$ and use the computational domain $(-6,6)\times(-6,6)$ with 300 cells in each 
direction.   Fig. \ref{stab} shows the initial configuration and the solution at $t=7$. 
One can see that the perturbation does not push the X-point away from its steady-state. 
On the contrary, the waves created by the perturbation leave the central area on the light 
crossing time and the steady-state configuration is restored. Although, here we present
the results only for this particular type of perturbation, we have tried several other types 
and obtained the same outcome.  Thus, we conclude that the magnetic X-point  
is stable to perturbations on a length scale which is much smaller compared to its size, even 
when the perturbation amplitude is substantial. 

%\lorenzo
{We have also verified the stability of unstressed X-points with PIC simulations. Here, no initial perturbation is imposed on the system. In the standard setup of anti-parallel reconnection, the system would grow unstable from particle noise. Here, we have demonstrated that an unstressed X-point is stable to noise-level fluctuations.}